\documentclass{elsart}

\usepackage{natbib}

\usepackage{natbib}

\usepackage{graphics}
 \usepackage{graphicx}
 \usepackage{epsfig}
\usepackage{amssymb}

\begin{document}

\begin{frontmatter}


\title{A WENO Algorithm for the Radiative Transfer and
    Ionized Sphere at Reionization}

\author[label1]{Jing-Mei Qiu},
\author[label1]{Chi-Wang Shu},
\author[label2,label3]{Long-Long Feng},
\author[label4]{Li-Zhi Fang}

\address[label1]{Division of
Applied Mathematics, Brown University, Providence, RI 02912, USA}
 \address[label2]{Purple Mountain Observatory, Nanjing, 210008,
P.R. China}
\address[label3]{National
Astronomical Observatories, Chinese Academy of Science, Chao-Yang
District, Beijing 100012, P.R. China}
\address[label4]{Department of Physics, University of Arizona,
Tucson, AZ 85721, USA}

\begin{abstract}

We show that the algorithm based on the weighted essentially
nonoscillatory (WENO) scheme with anti-diffusive flux corrections
can be used as a solver of the radiative transfer equations. This
algorithm is highly stable and robust for solving problems with
both discontinuities and smooth solution structures. We test this
code with the ionized sphere around point sources. It shows that
the WENO scheme can reveal the discontinuity of the radiative or
ionizing fronts as well as the evolution of photon frequency spectrum
with high accuracy on coarse meshes and for a very wide parameter space.
This method would be useful to study the details
of the ionized patch given by individual source in the epoch of
reionization. We demonstrate this method by calculating the
evolution of the ionized sphere around point sources in physical and
frequency spaces. It shows that the profile of the fraction of
neutral hydrogen and the ionized radius are sensitively dependent
on the intensity of the source.

\end{abstract}


\end{frontmatter}

\section{Introduction}

In the study of cosmic large scale structures formation, one
generally focuses on the inhomogeneity of the density and velocity
fields of dark matter and baryon gas, but assumes that the
distribution of background radiation is uniform. This would be a
reasonable approximation when the universe is transparent.
Therefore, the effect of radiative transfer is generally ignored in
the study of large scale structures on low redshift. However, the
fluctuations in the radiation field is essential in understanding
the early universe, especially in the epoches around reionization.
Radiative transfer problems are no longer to be overlooked. Many
numerical solvers for the radiative transfer equation have been
proposed (Abel et al. 1999, Ciardi et al. 2001, Gnedin \& Abel 2001,
Sokasian et al. 2001, Razoumov et al. 2002, Cen 2002, Maselli et al.
2003).

In this paper, we will introduce the algorithm of the radiative
transfer equations based on the finite difference weighted
essentially non-oscillatory (WENO) scheme (Jiang \& Shu 1996)
modified with anti-diffusive flux corrections (Xu \& Shu 2005). It
has been shown that the Boltzmann equations can be solved by this
WENO method with reasonable computational speed at least for one
or two spatial dimensions and two or three phase space dimensions
plus time (Carrillo et al. 2003, 2006). The WENO code is effective
to capture discontinuities as well as to resolve complicated flow
features. Moreover, it is also significantly superior over
piecewise smooth solutions containing discontinuities. It
has advantages over the Monte-Carlo codes in eliminating fluctuations,
in obtaining accurate probability density functions, and in
accurately resolving time transients of the system. Considering
the equation of radiative transfer basically is the same as
the Boltzmann equation, the WENO algorithm for the radiative transfer
is straightforward. Moreover, a hybrid algorithm of
hydrodynamic/N-body simulation based on WENO scheme has been
successfully developed (Feng et al. 2004). This result also
motivated us to extend the WENO scheme to radiative transfer
problems. It would be a necessary preparation to develop solver
of hydrodynamic/radiative transfer problems.

We will demonstrate the WENO algorithm by applying
the WENO code to the Str\"omgren sphere. The structure and
evolution of an individual Str\"omgren sphere are important for the
understanding of the reionization process. The ionization of baryon
gas surrounding high redshift objects, such as quasars, stars of first
generation, generally is modelled by the Str\"omgren sphere
(Madau \& Rees, 2000; Cen \& Haiman 2000; White et al. 2003; Stuart
et al. 2004;
Yu \& Lu 2005). The Str\"omgren sphere usually is described as
a fully ionized region, sharply divided from the neutral hydrogen
region on the outside of the sphere. Therefore, we should require
the algorithm to be able to 1.) capture the sharp boundary between
the ionized and neutral regions, and their time-dependence; 2.)
accurately calculate the small amount of neutral hydrogen in the
sphere and ionized hydrogen outside the sphere. These two points
can be realized by the the WENO numerical scheme, as it is effective
for problems containing both strong  discontinuities and smooth
flow structures.

The paper is organized as follows. Section 2 describes the WENO
numerical scheme. Section 3 gives the numerical solutions of the
Str\"omgren sphere.  The discussion and conclusion are presented
in \S 4. The derivation of the equation of the radiative transfer
is given in the Appendix.

\section{Numerical solver of the WENO scheme}

The radiative transfer equation in an expanding universe is (see
the Appendix)
\begin{equation}
\label{eq1}
        {\partial J\over\partial \, (ct)} +
        {\partial \over\partial x^i}\left (\frac{n^i}{a} J \right )
        +
        {\partial \over\partial \omega}(HJ) =
        - (k_\nu+3H) J + S,
\end{equation}
where $J(t, {\bf x}, \nu, n_i)$ is the specific intensity, $a$ the
cosmic factor, $H=\dot{a}/a$, $\nu$ the frequency of photon,
$\omega\equiv \ln 1/\nu$, and $n_i$ a unit vector in the direction
of photon propagation. $k_\nu$ and $S$ are, respectively, the
absorption and sources of photons.

To demonstrate the WENO method, we consider the simplest case. We
ignore the expansion of the universe. This approximation is
correct if the time scale of evolution of photon distribution is
much less than the time scale of cosmic expansion. Moreover, the
physical and phase spaces are assumed to be 1-dimensional with
coordinate $r$. In this case, the radiative transfer equation is
\begin{equation}
\label{eq2} {\partial J\over\partial ct} +
        \frac{\partial J}{\partial r}
          = - k_\nu J + S
\end{equation}
The computational domain is  discretized into a tensor product
mesh. The mesh is taken to be uniform in the $r$-direction, and to be
smooth non-uniform in the $\nu$-direction, i.e.
\begin{eqnarray}
\label{eq3}
r_i & = &  i \Delta r; \qquad i=0, 1, 2, ... , N_r ,
\nonumber \\
\nu_j & = & 2^{\xi_j}; \qquad j=0, 1, 2, ... , N_{\nu} ,
\end{eqnarray}
where $\Delta r = r_{\max} /N_r$ is the mesh size in the $r$-direction,
and $\xi_j= j \Delta \xi$, $\Delta \xi =
 {\log_2}{\nu_{\max}}/N_{\nu}$ is the transformed mesh size in the $\nu$
direction. $r_{\max}$ and $\nu_{\max}$ are, respectively, the sizes
of the numerical domain, which are adjusted in the numerical
experiment such that $J(t, r, \nu) \simeq 0$, for $r > r_{\max}$
and all $t$ and $\nu$, and $J(t, r, \nu) \simeq 0$, for $\nu >
\nu_{\max}$ and all $t$ and $r$.

The approximations to the point values of the solution $J(t^n,
r_i, \nu_j)$, denoted by ${J^n_{i, j}}$, are obtained with an
approximation to the spacial derivatives using the 5th order
finite difference WENO
method (Jiang \& Shu 1996) with anti-diffusive flux corrections
(Xu \& Shu 2005).

\begin{itemize}

\item Approximation to the derivatives:

\bigskip

To calculate $\partial J/\partial r$, the variable $\nu$
is fixed and the approximation is performed along the $r$-line
$$
{\partial \over {\partial r}}J(t^n, r_i, \nu_j) \approx
\frac{1}{\Delta r} \left( {\hat{h}^a}_{i+1/2} -
{\hat{h}^a}_{i-1/2} \right)
$$
where the numerical flux ${\hat{h}^a}_{i+1/2}$ is obtained with
the procedure given below. We can use the upwind fluxes without
flux splitting in the fifth order WENO approximation because the
wind direction is fixed (positive).
To obtain the sharp resolution of the contact
discontinuities, the anti-diffusive flux corrections
are used in our code.

First, we denote
$$
{h_i} = J(t^n, r_i, \nu_j), i=-2, -1, ..., {N_r}+2
$$
where $n$ and $j$ are fixed.
The numerical flux from the regular WENO procedure is obtained by
$$
\hat{h}_{i+1/2}^{-} = \omega_1 \hat{h}_{i+1/2}^{(1)}
+ \omega_2 \hat{h}_{i+1/2}^{(2)} + \omega_3 \hat{h}_{i+1/2}^{(3)}
$$
where $\hat{h}_{i+1/2}^{(m)}$ are the three third order fluxes on
three different stencils given by
\begin{eqnarray*}
\hat{h}_{i+1/2}^{(1)} & = &
\frac{1}{3} h_{i-2} - \frac{7}{6} h_{i-1}
               + \frac{11}{6} h_{i}, \\
\hat{h}_{i+1/2}^{(2)} & = &
-\frac{1}{6} h_{i-1} + \frac{5}{6} h_{i}
               + \frac{1}{3} h_{i+1}, \\
\hat{h}_{i+1/2}^{(3)} & = &
\frac{1}{3} h_{i} + \frac{5}{6} h_{i+1}
               - \frac{1}{6} h_{i+2},
\end{eqnarray*}
and the nonlinear weights $\omega_m$ are given by
$$
\omega_m = \frac {\tilde{\omega}_m}
{\sum_{l=1}^3 \tilde{\omega}_l},\qquad
 \tilde{\omega}_l = \frac {\gamma_l}{(\varepsilon + \beta_l)^2} ,
$$
with the linear weights $\gamma_l$ given by
$$
\gamma_1=\frac{1}{10}, \qquad \gamma_2=\frac{3}{5},
\qquad \gamma_3=\frac{3}{10},
$$
and the smoothness indicators $\beta_l$ given by
\begin{eqnarray*}
\beta_1 & = & \frac{13}{12} \left( h_{i-2} - 2 h_{i-1}
                             + h_{i} \right)^2 +
         \frac{1}{4} \left( h_{i-2} - 4 h_{i-1}
                             + 3 h_{i} \right)^2  \\
\beta_2 & = & \frac{13}{12} \left( h_{i-1} - 2 h_{i}
                             + h_{i+1} \right)^2 +
         \frac{1}{4} \left( h_{i-1}
                             -  h_{i+1} \right)^2  \\
\beta_3 & = & \frac{13}{12} \left( h_{i} - 2 h_{i+1}
                             + h_{i+2} \right)^2 +
         \frac{1}{4} \left( 3 h_{i} - 4 h_{i+1}
                             + h_{i+2} \right)^2 .
\end{eqnarray*}
$\varepsilon$ is a parameter to avoid the denominator
to become 0 and is taken as $\varepsilon = 10^{-5}$ times
the maximum magnitude of the initial condition $J$ in the
computation of this paper. The reconstruction of the finite difference
WENO flux on the downwind side $\hat{h}_{i+1/2}^{+}$ is obtained
in a mirror symmetric fashion with respect to $x_{i+1/2}$ as that
for $\hat{h}_{i+1/2}^{-}$.

The anti-diffusive flux corrections are based on the fluxes
obtained from the regular WENO procedure. It is given by
\begin{eqnarray}
\label{eq4} {\hat{h}^a}_{i+1/2} & = &
               \hat{h}_{i+1/2}^{-} + \\ \nonumber
 & &  \phi_{i}
{\rm minmod}\left (\frac{{h_i} - {h_{i-1}}}{\eta} + \hat{h}_{i-1/2}^{-} -
    \hat{h}_{i+1/2}^{-},
\hat{h}_{i+1/2}^{+}-\hat{h}_{i+1/2}^{-} \right ),
\end{eqnarray}
where $\eta= \triangle t/ \triangle r$ is the cfl number
and the minmod function is defined as
\begin{equation}
\label{eq5}
   {\rm minmod}(a,b)=\left\{\begin{array} {l}
        \displaystyle    0,\hspace{1cm} if\hspace{0.5cm} ab\leq 0 \\
        \displaystyle    a,\hspace{1cm}
  if\hspace{0.5cm} ab > 0, \mid a\mid \leq \mid b \mid \\
        \displaystyle    b,\hspace{1cm}
   if\hspace{0.5cm} ab > 0, \mid b\mid < \mid a \mid . \\
         \end{array}
   \right.
\end{equation}
$\phi_{i}$ of eq.(\ref{eq4}) is the discontinuity indicator
between 0 and 1, defined as
\begin{eqnarray*}
\phi_i &=& \frac{\beta_i}{\beta_i + \gamma_i},
\end{eqnarray*}
where
$$
\beta_i  = (\frac{\alpha_i}{\alpha_{i-1}}+
   \frac{\alpha_{i+1}}{\alpha_{i+2}})^2, \qquad
\gamma_i = \frac{\mid {u_{\max}} - {u_{\min}}\mid^2}{\alpha_i}, \qquad
\alpha_i = (\mid h_{i-1}- h_i \mid + \zeta)^2,
$$
with $\zeta$ being a small positive number taken as $10^{-6}$
in our computation. $u_{\max}$ and $u_{\min}$
are the maximum and minimum values of $h_i$ for all grid points.
With the definition above, we will have
$0 \leq \phi_i \leq 1$. $\phi_i = O(\Delta r ^2)$ in the smooth
regions and $\phi_ i$ is close to 1 near strong
discontinuities.
The purpose of the anti-diffusive flux corrections is to improve
the resolution of contact discontinuities without sacrificing
accuracy and stability of the original WENO scheme.

\bigskip

\item Approximation to the evolution with time

\bigskip

The time-derivative $\partial/\partial t$ in each step
$\Delta t$ is calculated by the third order TVD Runge-Kutta method
as
\begin{eqnarray*}
J^{(1)} & = & J^n + \Delta t L(J^n, t^n)  \\
J^{(2)} & = & J^n + \frac{1}{4} \Delta t  L(J^{n}) +
\frac{1}{4} \Delta t L(J^{(1)}) \\
J^{n+1} & = & J^n + \frac{1}{6} \Delta t L(J^{n}) +
  \frac{1}{6} \Delta t L(J^{(1)})+ \frac{2}{3}\Delta t L(J^{(2)})
\end{eqnarray*}
where $L$ is the approximation to the spatial derivatives and
the source terms:
$$
L(J) \approx
- \frac{\partial }{\partial r} J
- k_{\nu} J +S
$$

The Runge-Kutta method needs to be modified considering
the modification on the anti-diffusive flux $\hat{f}^a$ by
\begin{eqnarray*}
J^{(1)} & = & J^n + \Delta t L(J^n, t^n)  \\
J^{(2)} & = & J^n + \frac{1}{4} \Delta t  L'(J^{n}) +
\frac{1}{4} \Delta t L(J^{(1)}) \\
J^{n+1} & = & J^n + \frac{1}{6} \Delta t L''(J^{n}) +
  \frac{1}{6} \Delta t L(J^{(1)})+ \frac{2}{3}\Delta t L(J^{(2)})
\end{eqnarray*}
where the operator $L$ is defined by the anti-diffusive flux
$\hat{h}^a$ given by eq.(\ref{eq4}), and the operator
$L'$ is defined by the modified anti-diffusive flux
$\overline{h}^a$ as
\begin{equation}
\label{eq6}
   \overline{h}_{i+1/2}^a=\left\{\begin{array} {l}
        \displaystyle    \hat{h}_{i+1/2}^{-}+
  {\rm minmod}\left (\frac{4(h_i-h_{i-1})}{\eta} \right .\\
   \displaystyle \hspace{0.3cm}
            \left . +\hat{h}_{i-1/2}^{-}-\hat{h}_{i+1/2}^{-},
                         \hat{h}_{i+1/2}^{+}-\hat{h}_{i+1/2}^{-}\right ),
                     \\
    \displaystyle \hspace{3cm} if\hspace{0.1cm} bc>0, \mid b\mid
         <\mid c\mid, \\
   \displaystyle    \hat{h}_{i+1/2}^a,\hspace{2cm} {\rm otherwise}\\
      \end{array}
   \right.
\end{equation}
and $L''$ is defined by the modified anti-diffusive flux
$\widetilde{h}^a$,
\begin{equation}
\label{eq7} \widetilde{h}_{i+1/2}^a=\left\{\begin{array} {l}
        \displaystyle    \hat{h}_{i+1/2}^{-}
      + {\rm minmod}\left (\frac{6(h_i-h_{i-1})}{\eta} \right .\\
\displaystyle  \hspace{0.3cm}
      \left . +\hat{h}_{i-1/2}^{-}-\hat{h}_{i+1/2}^{-},
                         \hat{h}_{i+1/2}^{+}-\hat{h}_{i+1/2}^{-} \right ), \\
\displaystyle \hspace{3cm} if\hspace{0.5cm} bc>0,
     \mid b\mid <\mid c\mid, \\
        \displaystyle    \hat{h}_{i+1/2}^a,\hspace{1cm} {\rm otherwise}\\
         \end{array}
   \right.
\end{equation}
Here $b =(h_i-h_{i-1})/\eta +
{\hat{h}_{i-1/2}^{-}} - \hat{h}_{i+1/2}^{-}$, $c = {\hat{h}_{i+1/2}^{+}} -
\hat{h}_{i+1/2}^{-}$.
\end{itemize}

We have given the details of the WENO algorithm with
anti-diffusive flux corrections only for the one dimensional case
to save space. The finite difference WENO scheme is ideally suited
for multi-dimensional calculations, as derivatives in each
direction can be approximated in an one dimensional setting by
fixing the other variables, while still maintaining high order
accuracy and stability.  We refer to (Carrillo et al. 2003, 2006)
for more details of WENO approximations to Boltzmann equations.
Our radiative transfer equations [Eq.(\ref{eq1})] are of the same
form as the Boltzmann equations in (Carrillo et al. 2003, 2006)
and hence the algorithms developed there can be applied here
without difficulty.

\section{Ionized sphere}

\subsection{Ionized source in a uniform medium}

As a test, we consider a point photon source located at the center
${\bf x}=0$ of a uniformly distributed medium. Assuming the source is
monochromatic with frequency $\nu_0$, and all photons are emitted
along the radial direction, we have
\begin{equation}
\label{eq8} S(t, {\bf x},{\nu}, {\bf n})=f(t, {\bf x})\delta(\nu-\nu_0)
\delta({\bf n - e_r}).
\end{equation}
and
\begin{equation}
\label{eq9}
      f(t, {\bf x})=\left \{ \begin{array}{ll}
      E/V, & {\rm at \ the \ center \ {\bf x}=0} \\
      0,  & {\rm otherwise},
      \end{array} \right .
\end{equation}
where $E$ is the total energy of photons emitted from the sources
per unit time. When $V\rightarrow 0$, $f(t, {\bf x})\rightarrow E
\delta ({\bf x})$.

\begin{figure}[htb]
\centerline{
\includegraphics[width=2.0in,height=2.0in]{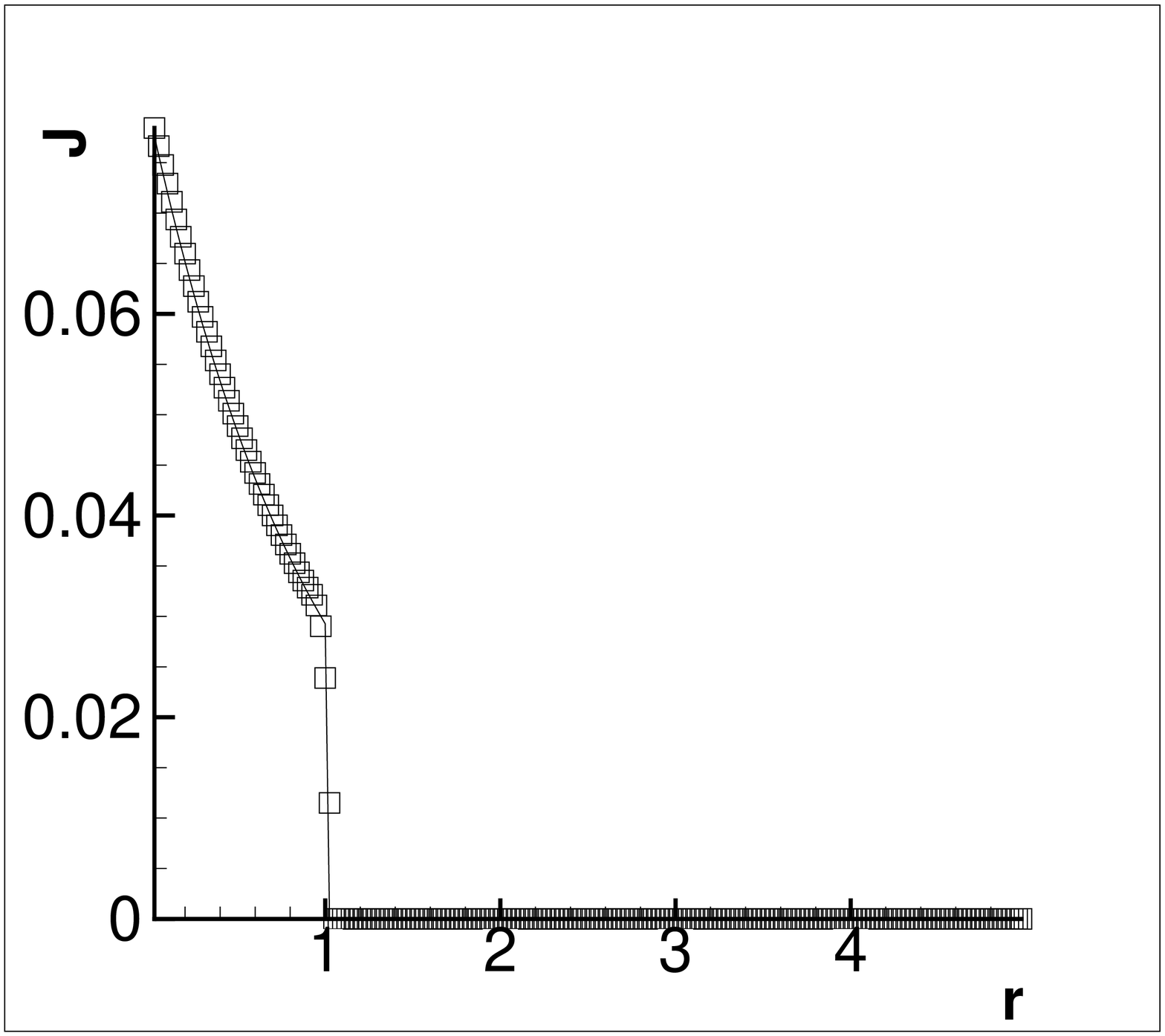}
\includegraphics[width=2.0in,height=2.0in]{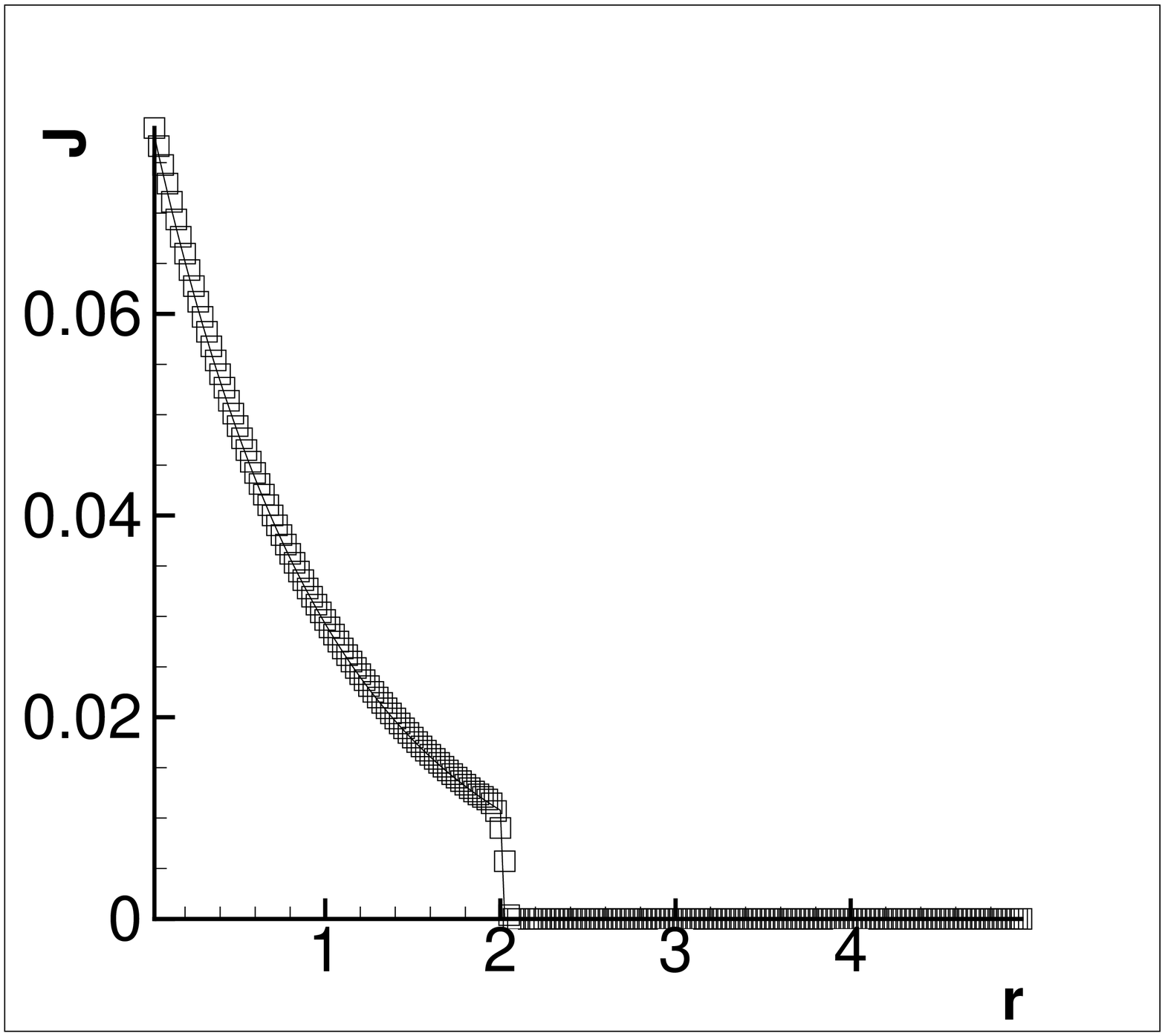}
}
\centerline{
\includegraphics[width=2.0in,height=2.0in]{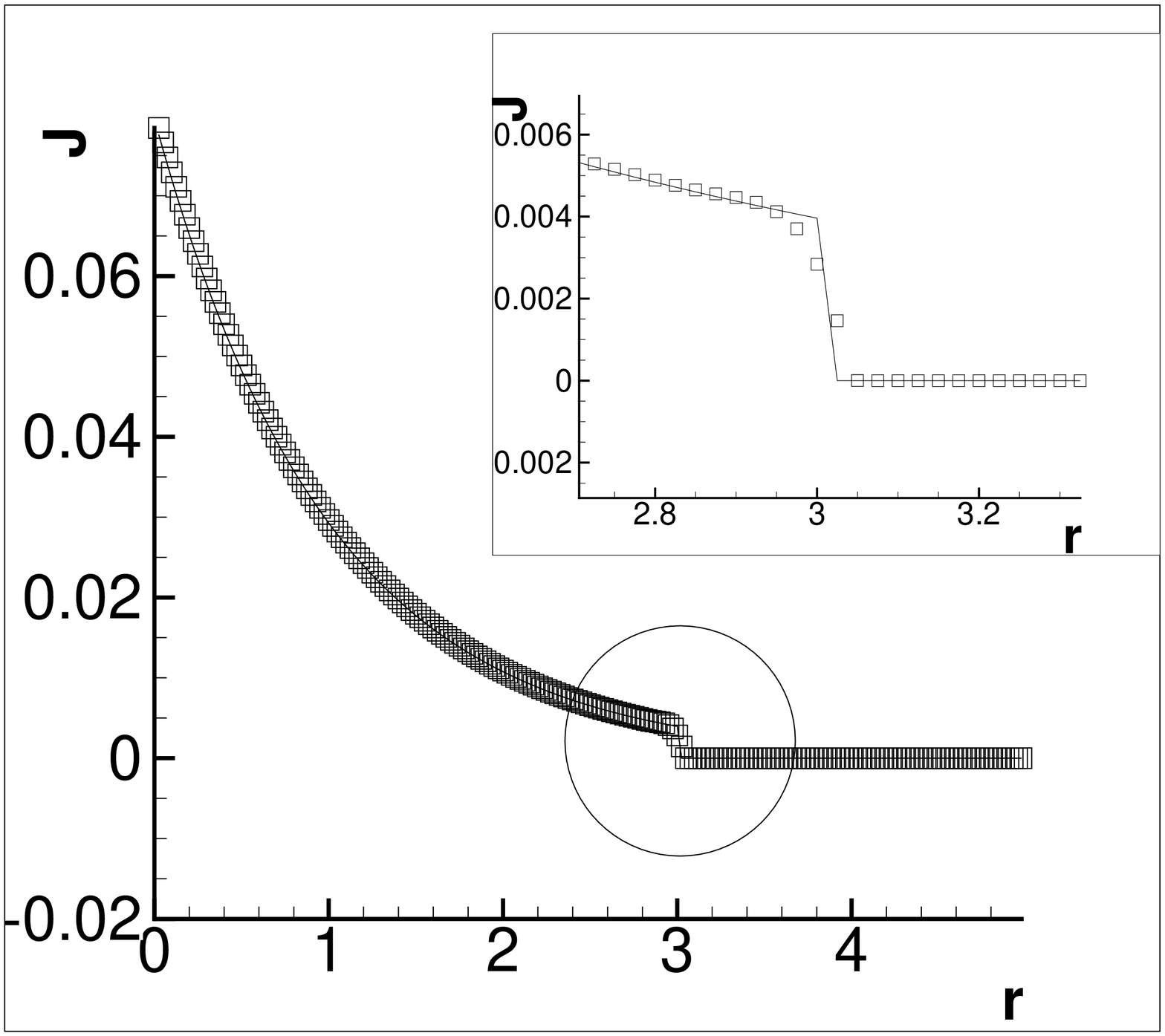}
\includegraphics[width=2.0in,height=2.0in]{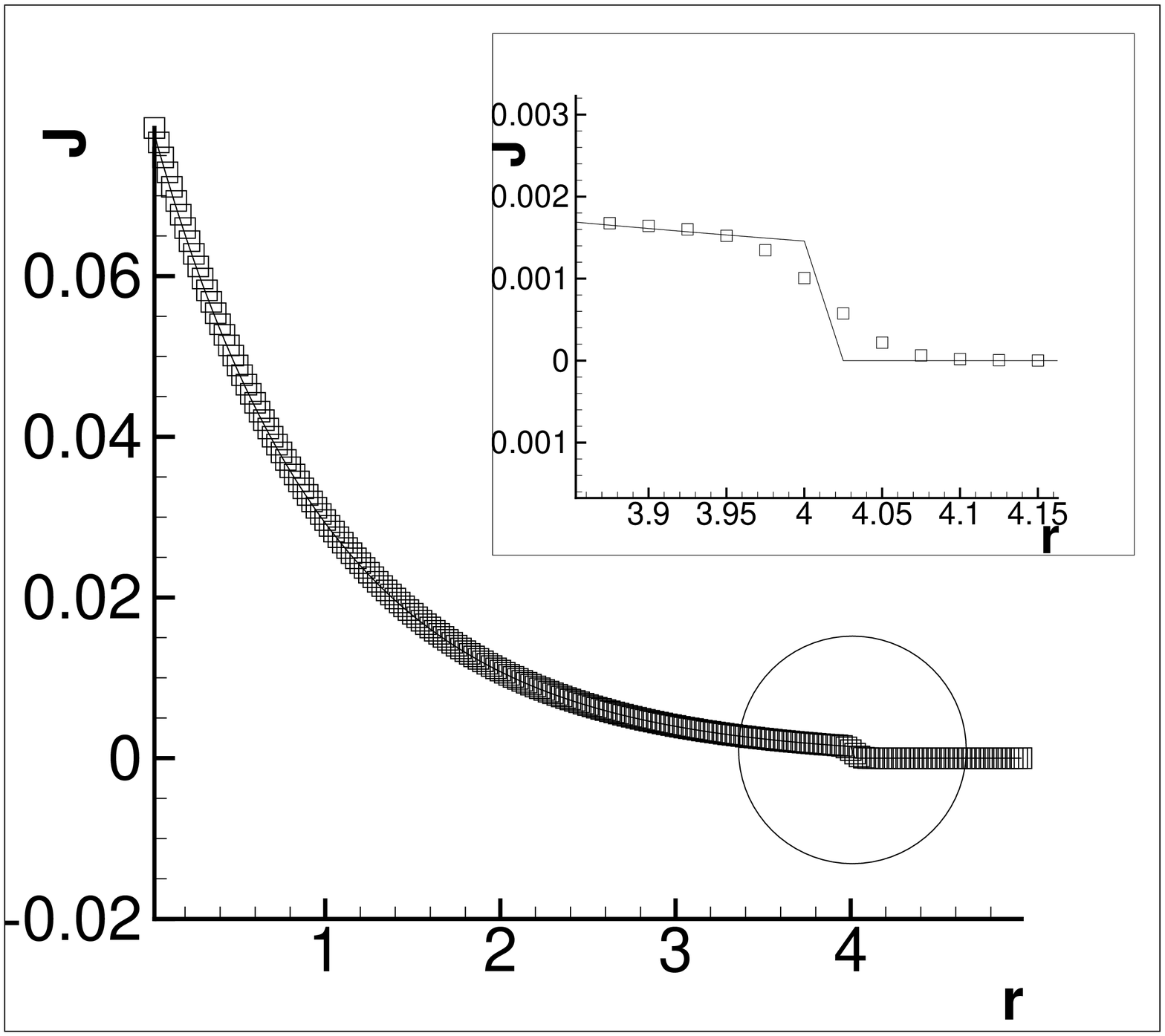}
}
\bigskip
\caption{The curves of $J(t,r)$ given by the numerical solution at
$N_r=200$
 (square) and the exact solution (solid line).
J(t=0, r)=$\frac{1}{4\pi}$.
The time is taken to be $t=1.0$ (top left); $t=2.0$ (top right);
$t=3.0$ (bottom left) and $t=4.0$ (bottom right).
The small windows in the bottom figures are the zoom-in version of the
corresponding regions in the circles.
}
\label{fig1}
\end{figure}

In this case, Eq.(\ref{eq1}) could be written in the spherical
coordinates as
\begin{equation}
\label{eq10}
        {\partial J\over\partial \, (ct)} + \frac {1}{r^2}
        \frac{\partial}{\partial r}\left (r^2 n^{r}J \right )
          = - k J, \hspace{5mm} {\rm r\neq 0}
\end{equation}
where the absorption coefficient $k$ is given by $k= n\sigma$, $n$
being the number density of particle of the medium, and $\sigma$
the absorption cross-section. In this problem, both $n$ and
$\sigma$ are assumed to be constants.

We define $r^2J \equiv J'(t,r)\delta(\nu-\nu_o)\delta({\bf n -
e_r})$. Then eq.(\ref{eq10}) becomes
\begin{equation}
\label{eq11} {\partial J'\over\partial \, (ct)} +
        \frac{\partial}{\partial r} J'
          = - kJ',  \hspace{5mm} {\rm r\neq 0}.
\end{equation}
For simplicity, we drop the prime, and use $J(t,r)$ for $J'(t,r)$
below. It will not cause confusion between the $J$ in
eqs.(\ref{eq11}) and (\ref{eq10}), as the former is a function of
$t$, $r$, while the later is a function of $t$, $r$, $\nu$ and
$n_i$. The source term gives a boundary condition at $r=0$ as
\begin{equation}
\label{eq12}
\lim_{r\rightarrow 0} 4\pi J(t,r) = E
\end{equation}
Assuming the source starts to emit photon at $t=0$, the initial
condition is then
\begin{equation}
\label{eq13} J(t=0, r)=\left \{\begin{array}{ll}
                               E/4\pi, & r=0\\
                                0,  & r>0.
                              \end{array}
                              \right .
\end{equation}

Subject to the conditions (\ref{eq12}) and (\ref{eq13}), the exact
solution of eq.(\ref{eq11}) is
\begin{equation}
\label{eq14} J(t,r) =\theta(t-r)\frac{E}{4\pi} e^{-kr},
\end{equation}
where $\theta(x)$ is a step function: $ \theta(x)=1$, if $x \geq
0$; and $\theta(x)=0$, if $x<0$. The step function is from the
radiative front, which propagates with the speed of light $r=t$.

Eq.(\ref{eq11}) has the same shape as Eq.(\ref{eq2}). Similarly, we
can apply the WENO scheme to solve Eq.(\ref{eq11}). The boundary
condition eq.(\ref{eq12}) makes us to take the inflow condition at
$r=0$ as
$$
J(t^n,r_{i})= E/{4 \pi}, \qquad for \qquad i = 0, -1, -2.
$$
The boundary condition at $r=r_{\max}$ is:
$$
J_{N_r+i, j} = J_{N_r-1, j} ,\qquad for \qquad i=0, 1, 2
$$

Figure \ref{fig1} plots both the numerical solution and exact
solution at time $t= 1.0$, 2.0, 3.0, 4.0. Clearly, the numerical
result displays an excellent agreement with the exact solution even
when the number $N_r$ is only 200. The discontinuity at the
radiative front $r=t$ is also well reproduced in the WENO scheme.

\subsection{The profile of an ionized sphere}

We now consider a realistic problem: the ionization of a uniformly
distributed hydrogen gas with number density $n$ by a point UV
photon source located at $r=0$. This problem is the well known
Str\"omgren sphere if the profile of the ionized sphere is
approximated by two sharply divided regions as
\begin{equation}
\label{eq15}
 f_{\rm HI}(r) \simeq \left \{\begin{array}{ll}
                       0 & {\rm if \ } r<R_s \\
                       1 & {\rm if \ }  r>R_s \\
                      \end{array}
                      \right .
\end{equation}
where $f_{\rm HI}\equiv n_{\rm HI}/n$, $n_{\rm HI}(t, {\bf x})$ is
the number density of neutral hydrogen, HI. Eq.(\ref{eq15}) means
that within the sphere of radius $R_s$ around the source, hydrogen
is fully ionized, while outside the sphere hydrogen atoms remain
neutral. $R_s$ is called the ionized radius of the Str\"omgren
sphere, which can be determined by the balance between the rate of
recombinations and the emission $\dot{N}$ of ionizing photons
(Str\"omgren 1939, Spitzer 1978, Osterbrook 1989)
\begin{equation}
\label{eq16} R_s=\left (\frac {3 \dot{N}}{4\pi \alpha_{\rm
HII}n^2}\right)^{1/3},
\end{equation}
where $\alpha_{\rm HII}$ is the hydrogen recombination
coefficient.

The sharp profile eq.(\ref{eq15}), and then eq.(\ref{eq16}), are
reasonable if the mean free path of photon $\lambda \simeq
1/\sigma_0 n$ is much less than $R_s$. For strong sources, like
quasars $\dot{N}\simeq 10^{57}$ s$^{-1}$, $R_s$ is much larger than
$\lambda$. However, for weak sources, like population III stars
$\dot{N}\leq 10^{50}$, $\lambda$ is comparable, or even larger than
$R_s$. In this case, the profile eq.(\ref{eq15}) is no longer valid.
Even for strong sources, we need to study the small deviation of
the profile of ionized sphere from eq.(\ref{eq15}). The profile
should be found by solving the
radiative transfer equation with proper boundary and initial
conditions. Moreover, the profile eq.(\ref{eq15}) describes the
final state of the ionized sphere around a strong source. To study
the reionization history of the universe, we may need the
information of the formation of the ionized sphere i.e. the
time-dependence of $f_{\rm HII}(t,r)$. This also requires to solve
the radiative transfer equation.

To calculate the profile of the ionized sphere, we can still use
eq.(\ref{eq11}), boundary condition eq.(\ref{eq12}), and initial
condition eq.(\ref{eq13}). The absorption coefficient now
is given by
\begin{equation}
\label{eq17}
 k_{\nu_0} = \sigma(\nu_0)n_{{\rm HI}}(t, {\bf x})
\end{equation}
where $\sigma(\nu_0)$ is the absorption cross section at the
ionization frequency $\nu_0$. We have $\sigma_0=\sigma(\nu_0)=6.3\times
10^{-18}$ cm$^2$.

The number density of neutral hydrogen $n_{\rm HI}(t, {\bf x})$ is
determined by the ionization equilibrium equation
\begin{equation}
\label{eq18}
\frac{df_{\rm HI}}{dt}= \alpha_{\rm HII}n_ef_{\rm HII} -
\Gamma_{\rm \gamma HI}f_{\rm HI}-\Gamma_{\rm eHI} n_ef_{\rm HI}.
\end{equation}
where $f_{\rm HI}(t,r)\equiv n_{\rm HI}/ n$,
$f_{\rm HII}(t,r)\equiv n_{\rm HII}/n$. We will assume electron density 
$n_e=n_{\rm HII}$. This means that
the electrons from ionized helium are ignored. The photoionization
is given by
\begin{equation}
\label{eq19}
{\Gamma_{\gamma {\rm HI}}}={1\over{r^2}}
{{J(t,r)}\over{h{\nu_0}}}\sigma_0 .
\end{equation}
The parameters $\alpha_{\rm HII}$ and $\Gamma_{\rm eHI}$ are the
recombination coefficient and collision ionization, respectively,
which can be found in e.g. Theuns et al (1998).

Let us rescale the variables by $t'=c\sigma(\nu_0) n t$,
$r'=\sigma(\nu_0) n r$. It means that $t'$ and $r'$ are,
respectively, the time and distance in the units of mean free
flight time and mean free path, $1/\sigma_0(\nu_0 )n$, of photon
$h\nu_0$ in neutral hydrogen gas with density $n$. In the
$\Lambda$CDM model, $n=1.88\times 10^{-7}(1+z_r)^3$ cm$^{-3}$,
where $z_r$ is the redshift of reionization, the unit of $t'$ is
0.89 $ \times 10^6(1+z_r)^{-3}$ years, and the unit of $r'$ is
0.27$(1+z_r)^{-3}$ Mpc. We also rescale the intensity by
$J''=J(\sigma^2 n/c h \nu_0)$. Eqs.(\ref{eq11}) can then be
rewritten as
\begin{equation}
\label{eq20} {{\partial J''}\over{\partial t'}}+
   {{\partial J''}\over{\partial r'}}=-f_{HI} J'', \hspace{5mm}
    r \ne 0
\end{equation}
The boundary condition (\ref{eq12}) and initial condition
(\ref{eq13}) are, respectively
\begin{equation}
\label{eq21}
J''(t,r=0)={J''_0}
\end{equation}
\begin{equation}
\label{eq22}
         J''(t=0,r)= \left \{\begin{array} {l}
        \displaystyle    0, \hspace{2cm} {\rm if}\hspace{0.5cm} r>0\\
        \displaystyle    J''_0,  \hspace{1.8cm}
        {\rm if}\hspace{0.5cm} r=0
         \end{array}
   \right .
\end{equation}
With this rescaling, the photoionization term of eq.(\ref{eq19}) is
$\Gamma_{\rm \gamma HI}/n=(J''/r'^2)(cn\sigma_0)$. The intensity of
source is
$\dot{N}=4\pi J(0)/h\nu=5.05 \times 10^{52}
J''_0(1+z_r)^{-3}$ s$^{-1}$.

\subsubsection{Strong sources}

\begin{figure}[htb]
\centerline{
\includegraphics[width=2.0in,height=1.5in]{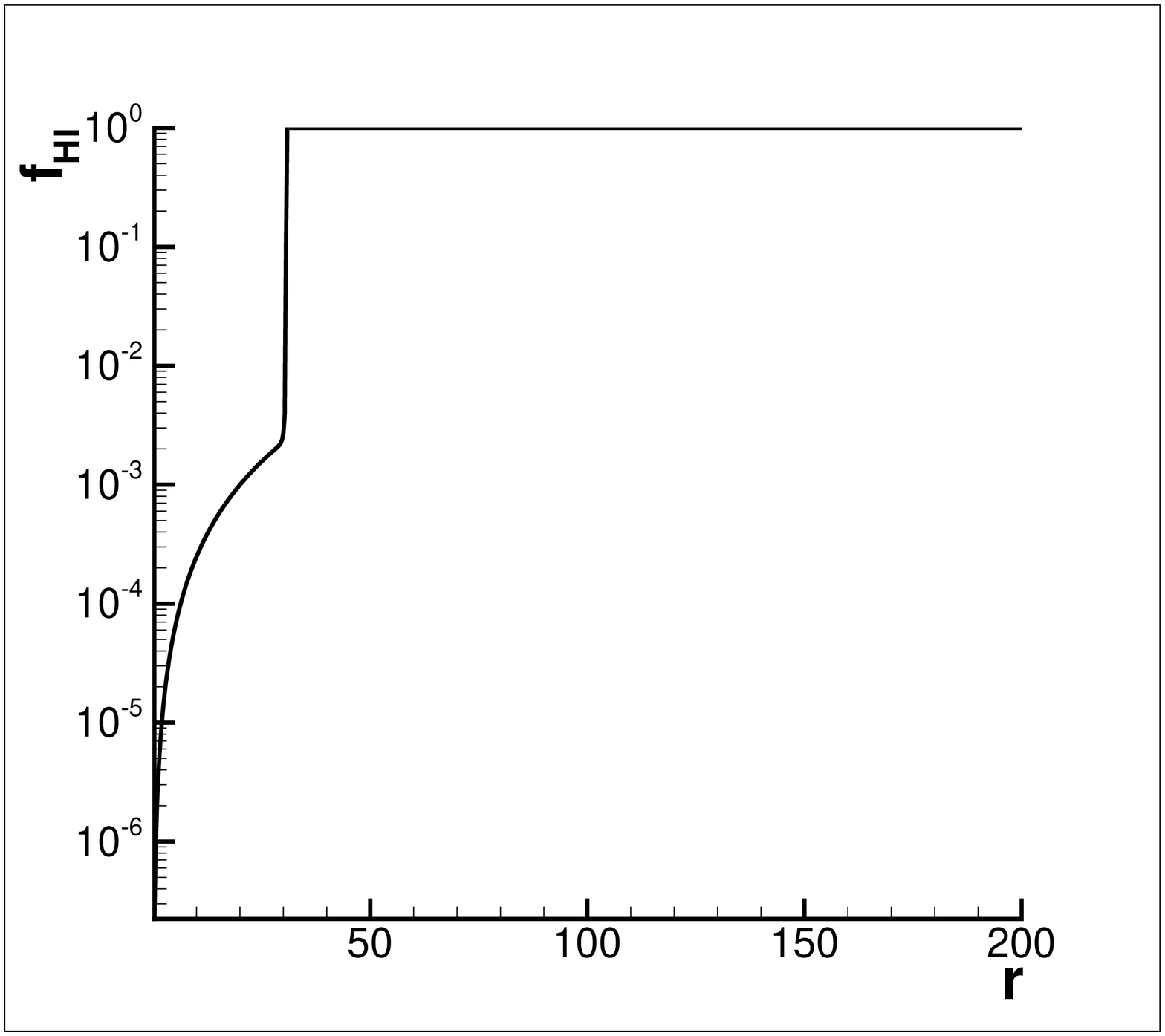}
\includegraphics[width=2.0in,height=1.5in]{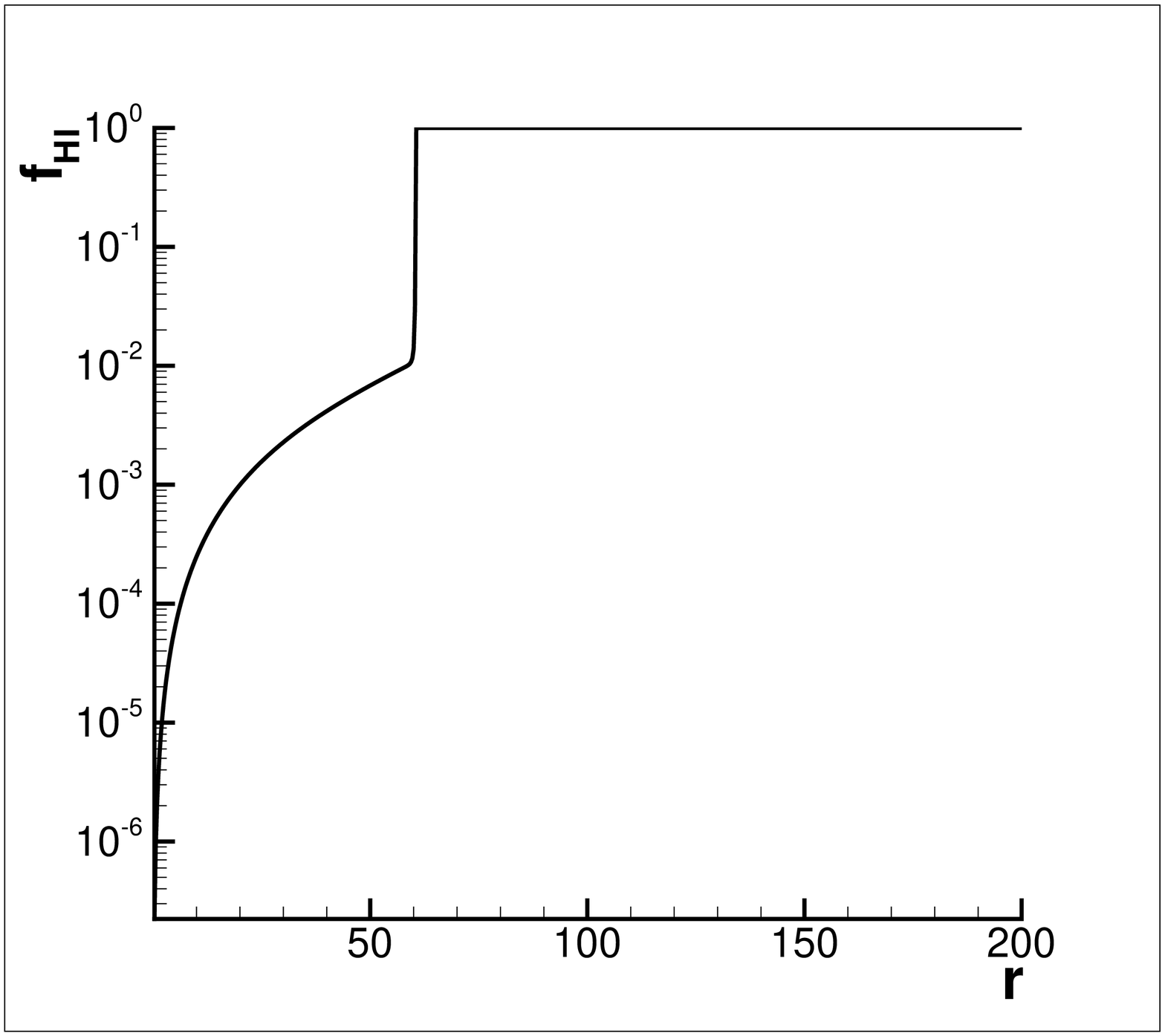}
}
\centerline{
\includegraphics[width=2.0in,height=1.5in]{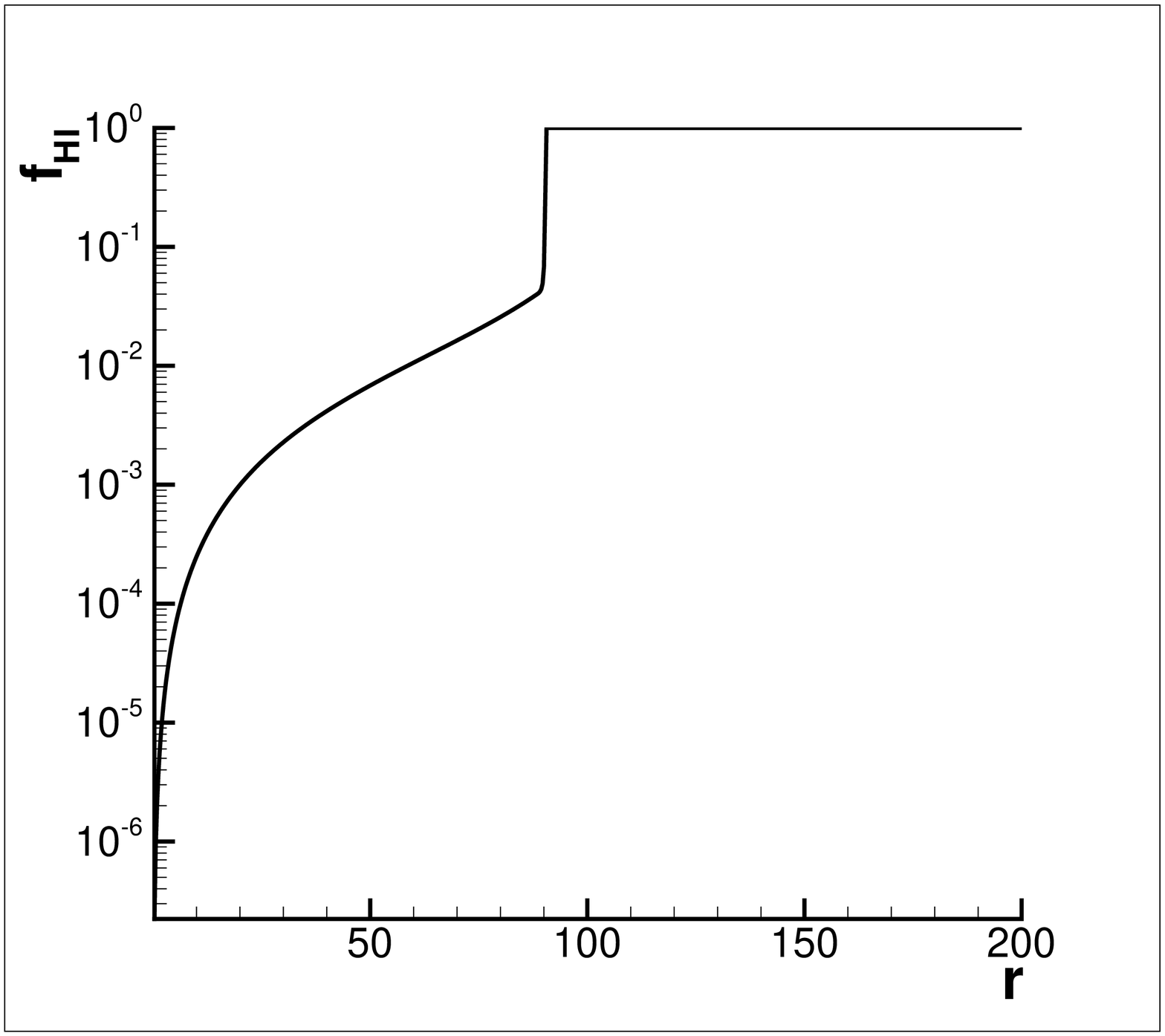}
\includegraphics[width=2.0in,height=1.5in]{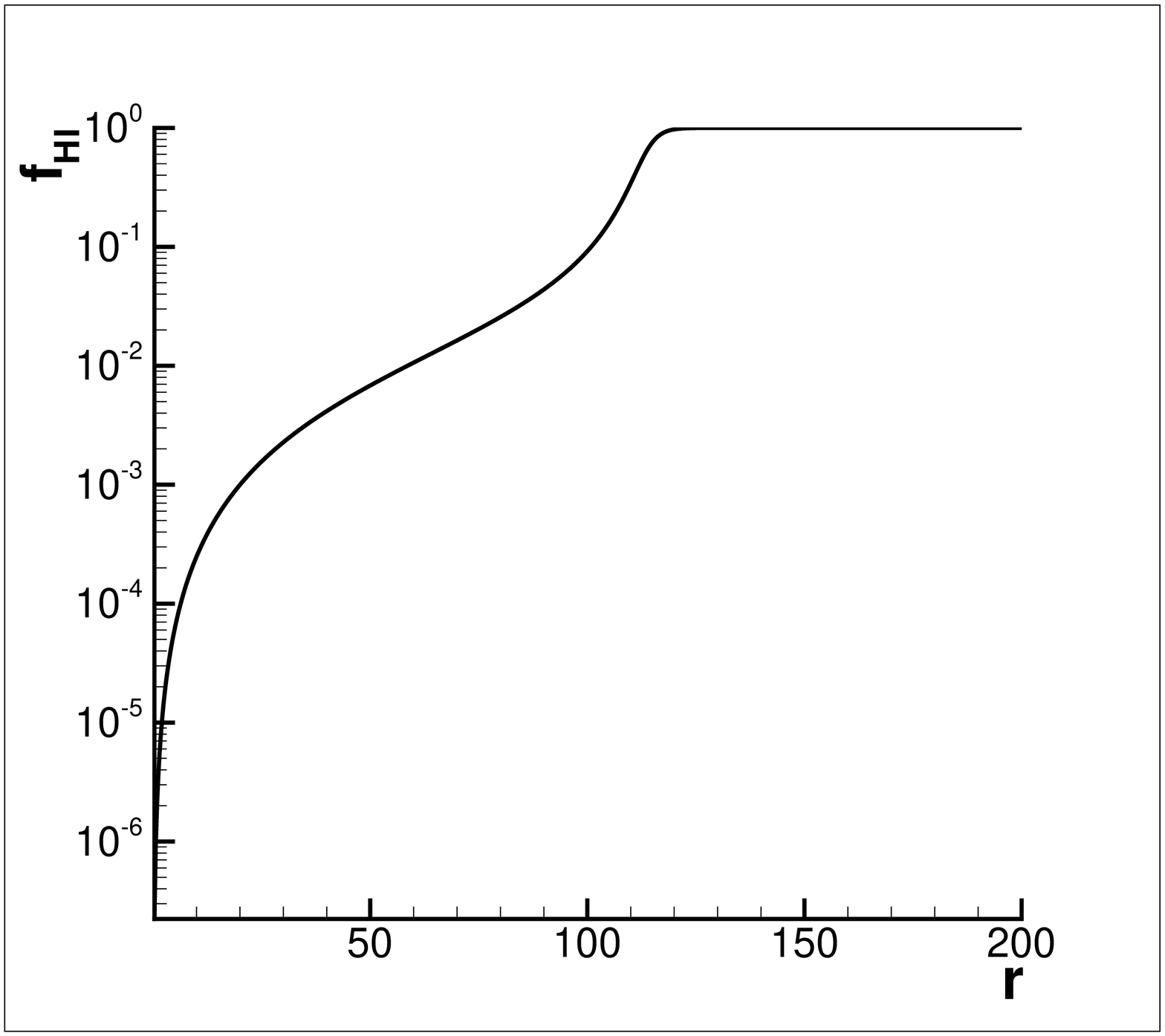}
}
\caption{
$f_{\rm HI}(t,r)$ vs. $r$ at time $t=30.0$ (top
left), 60.0 (top right), 90.0 (bottom left), 140.0 (bottom right)
for a strong source $J''_0=4.2$, or
$\dot{N}=2.16\times 10^{53}(1+z_r)^{-3}$ s$^{-1}$. The parameter
$N_r$ of
the numerical
calculation is taken to be 1000. The ionized radius $R_s$=122.04.}
\label{fig2}
\end{figure}

For strong sources, $R_s$ is much larger than the mean free path. We can
replace the rate equation (18) by ionization equilibrium equation
$df_{\rm HI}/dt=0$, or $\alpha_{\rm HII}f^2_{\rm HI}-
(\Gamma_{\gamma, {\rm HI}}/n +2\alpha_{\rm HII})f_{HI}+\alpha_{\rm HII}=0$, 
where the small term $\Gamma_{e {\rm HI}}$ is also dropped. A typical 
numerical result of the
$f_{\rm HI}(t,r)$ profile is shown in Figure \ref{fig2}, in which $J''_0=4.2$,
or $\dot{N}=2.16\times 10^{53}(1+z_r)^{-3}$. The calculation is performed with
$N_r=1000$.

Figure \ref{fig2} shows that the code can well reveal the jump at
radiative front $r=t$, which propagates with the speed of light.
The evolution of the ionized range can approximately be described as
\begin{equation}
\label{eq23}
f_{\rm HI}(t,r)\simeq \left \{ \begin{array}{ll}
                         \theta(r-t), & t \leq t_s\\
                         \theta(r-t_s), & t  >  t_s\\
                         \end{array}
                         \right .
\end{equation}
When $t$ is small, the
increase of ionized sphere is following the radiative front $r=t$.
The increase will be halted, when the ionization equilibrium is
totally established, and the ionized sphere becomes
time-independent. For the case of $J''(0)=4.2$, the final state
arrives at $t_s \simeq 1.22\times 10^2$. Since the bottom right panel
of Figure
\ref{fig2} is for $t=140  > t_s$, it should be the final
state of the ionized sphere. The profile of the final state can be
approximately described by the Str\"omgren sphere profile
eq.(\ref{eq15}), and ionized radius $R_s$ is about 122 mean free path.

\begin{figure}[htb]
\centerline{
\includegraphics[width=2.0in,height=1.5in]{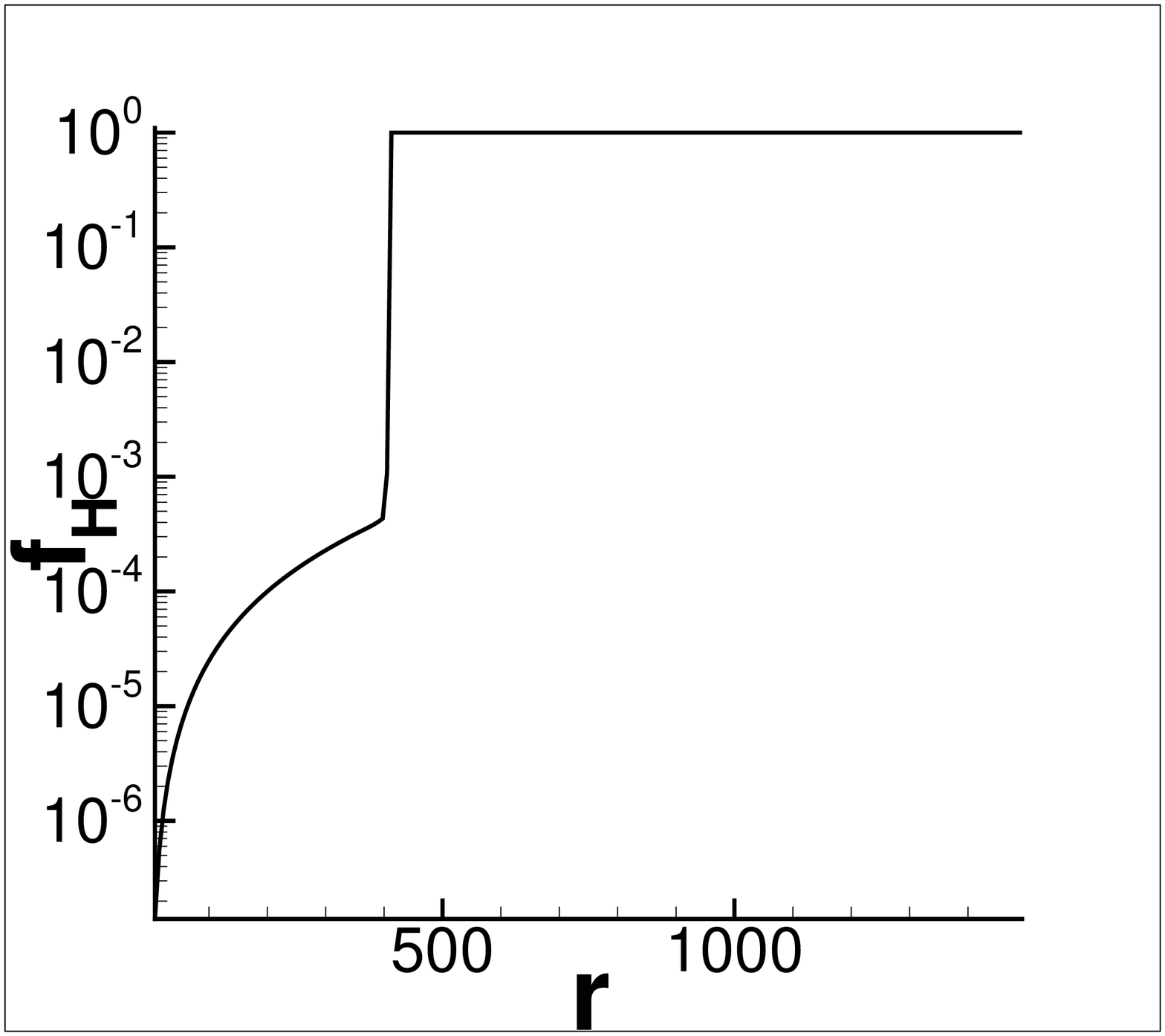}
\includegraphics[width=2.0in,height=1.5in]{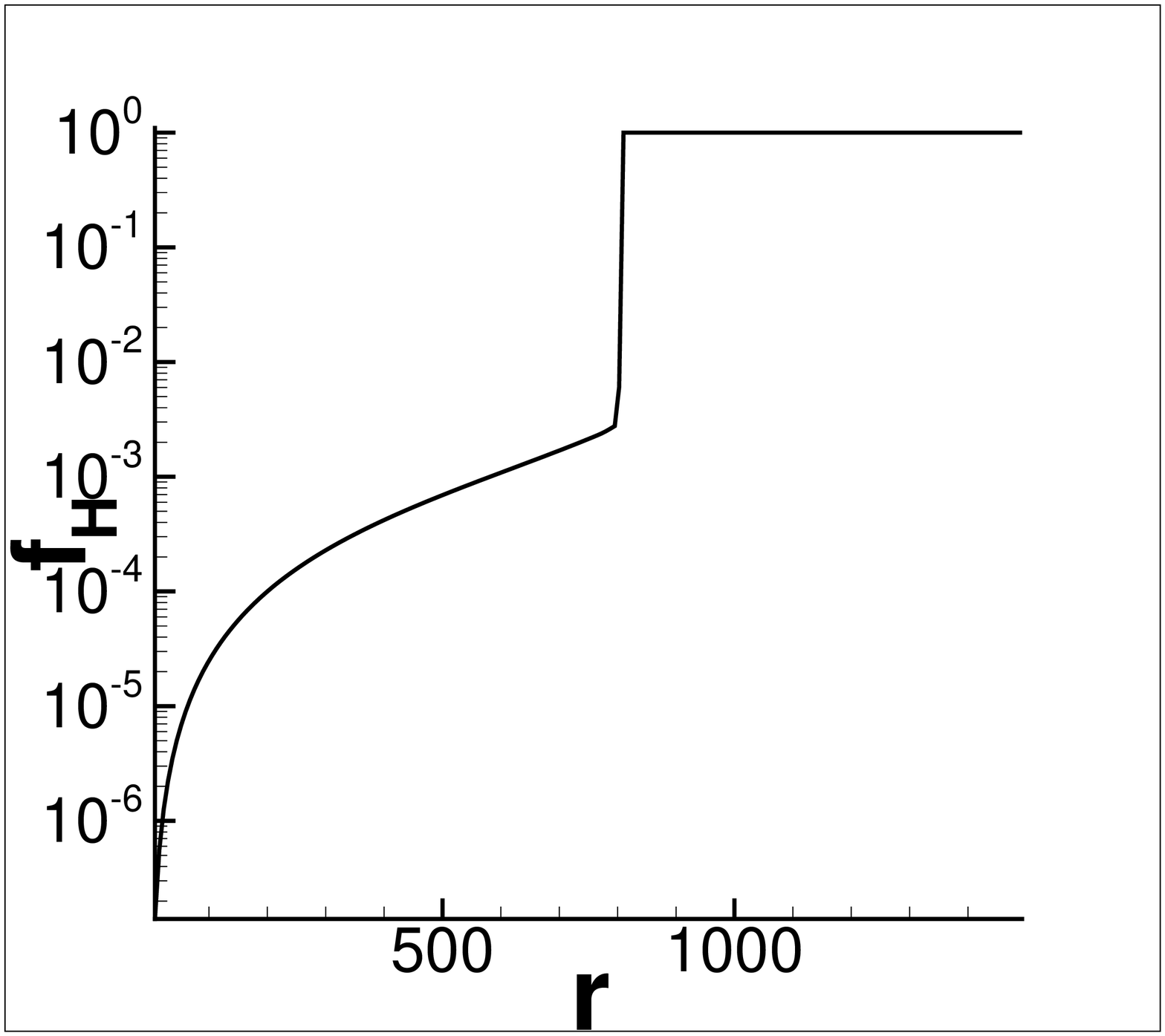}
} \centerline{
\includegraphics[width=2.0in,height=1.5in]{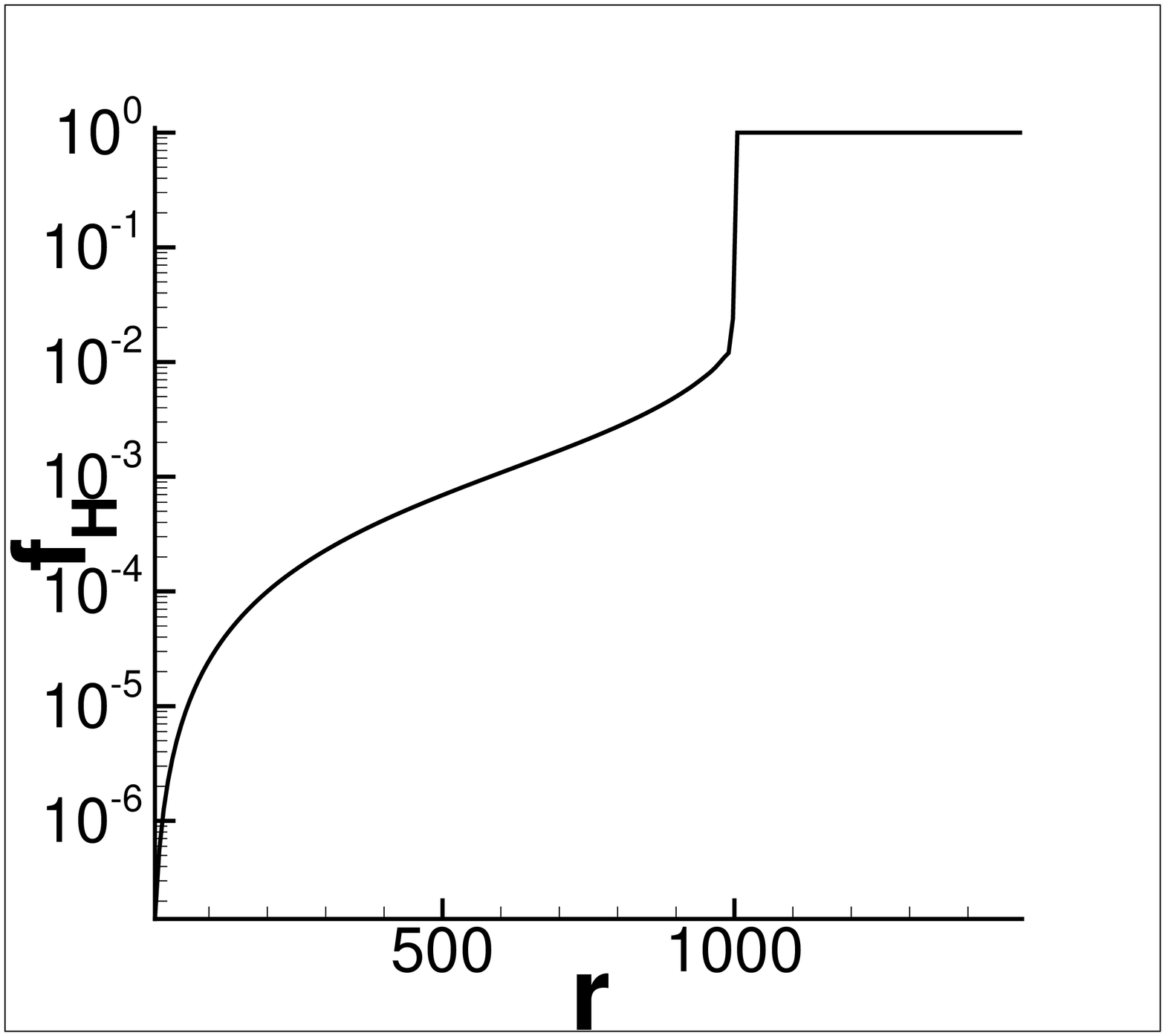}
\includegraphics[width=2.0in,height=1.5in]{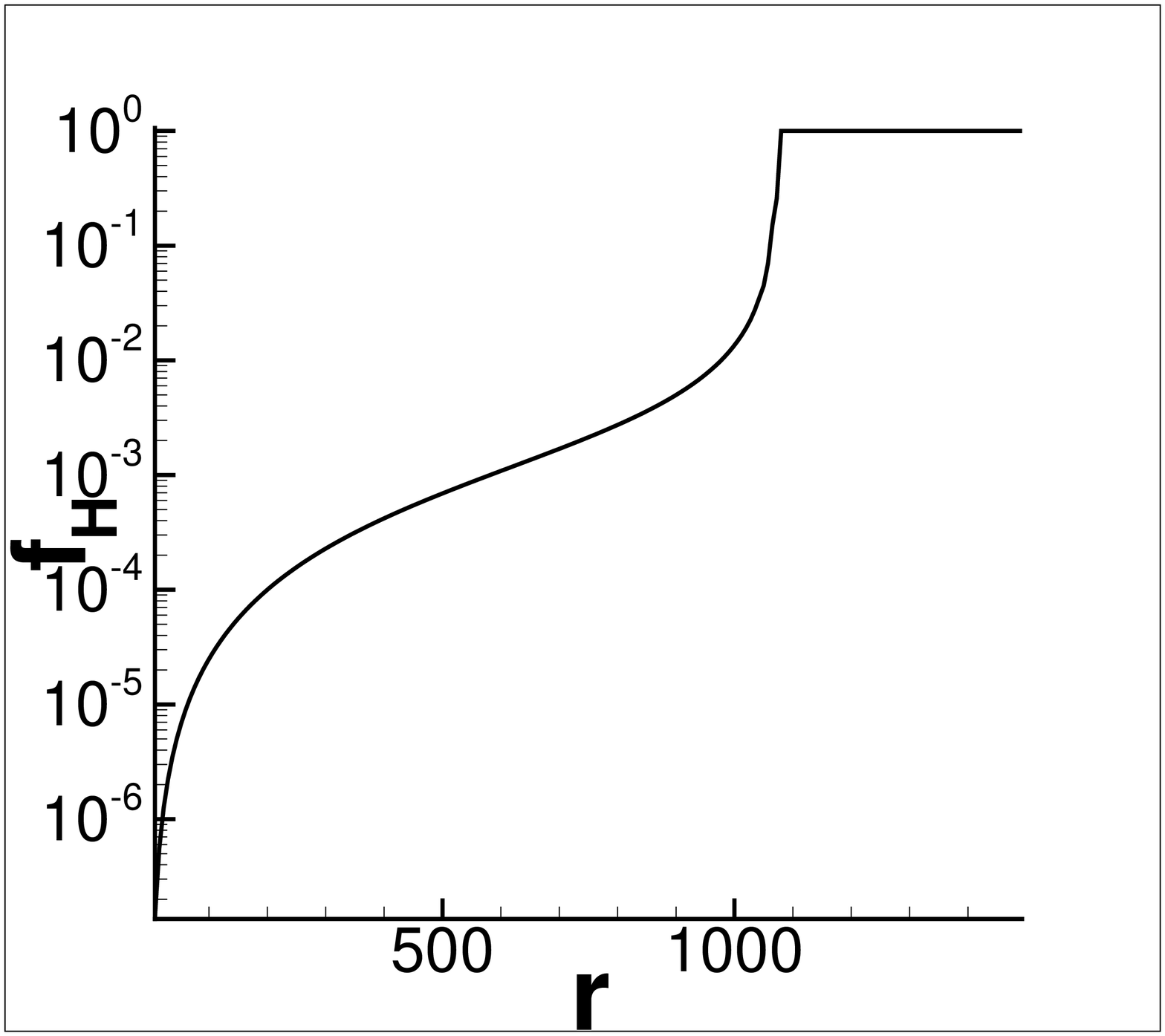}
}
\caption{$ f_{\rm HI}(t,r)$ vs. $r$ at time $t=400.0$ (top
left), 800.0 (top right), 1000.0 (bottom left), 1100.0 (bottom right)
for a strong source ${J_0}''=4.2\times 10^3$ or
$\dot{N}=2.6\times 10^{56}(1+z_r)^{-3}$ s$^{-1}$.
The parameter $N_r$ of the numerical calculation is taken to be 200.
The ionized radius $R_s$=1080.0.
}
\label{fig3}
\end{figure}

If we define an effective ionized radius $r_{\rm HII}(t)$ by
\begin{equation}
\label{eq24}
\frac{1}{3}r_{\rm HII}^3(t)=\int_{0}^{r=t}f_{\rm HII}(r,t)r^2dr,
\end{equation}
we have $r_{\rm HII}(t)\simeq t$ when $t < t_s$. The physical
meaning of $r_{\rm HII}$ is clear. It gives an equivalent sphere,
within which hydrogen is fully ionized. The ionized radius is
essentially the same as the radiative front. Therefore, for strong
sources, the radiative front and ionized radius are the same in all
time $t<t_s$.

 From Figure \ref{fig2} we can also see that the neutral hydrogen
fraction within the ionized sphere $r<t_s$ actually is not exactly
equal to zero. Although the profile (\ref{eq15}) may still be
reasonable, $f_{\rm HI}$ can be as large as 10$^{-2}$ near the
ionized radius. We also calculated the profile with very strong
source like $J''_0=4.2\times 10^3$, which is shown in Figure \ref{fig3}. 
In this case
the source intensity $J''_0$ is larger than that of Figure \ref{fig2} by
a factor of 10$^3$, and therefore, the radius of $R_s$ of Figure \ref{fig3}
is larger than that of \ref{fig2} by a factor of 10. There is also 
neutral hydrogen left behind the ionization front. These
small amounts of HI are stable, i.e. independent of the parameter $N_r$.
The neutral hydrogen remained in the ionized sphere is not always
negligible.

\subsubsection{Weak Sources}

If the Str\"omgren sphere radius $R_s$ of a source is comparable or
less than the mean free path, it is a weak sources. In this case, we
should use eq.(18) to describe the ionization evolution. In the range of
$r$ less than the mean free path, i.e. $r'\ll 1$, the photonionization
$\Gamma_{\rm HI}$ is large. We can keep only the photoionization
term $\Gamma_{\rm HI}f_{\rm HI}$ on the r.h.s. of eq.(18).

\begin{figure}[htb]
\centerline{
\includegraphics[width=2.0in,height=1.5in]{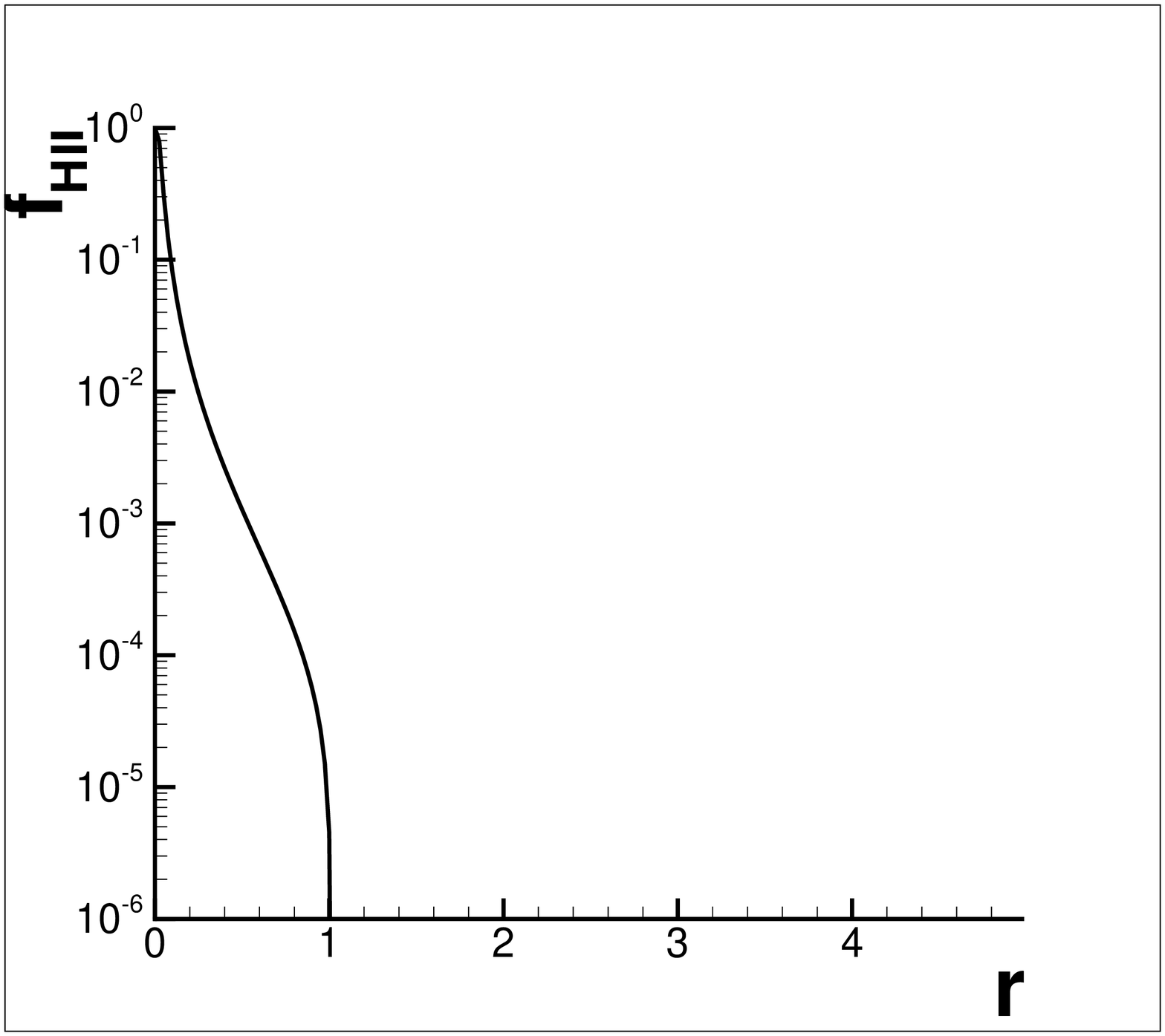}
\includegraphics[width=2.0in,height=1.5in]{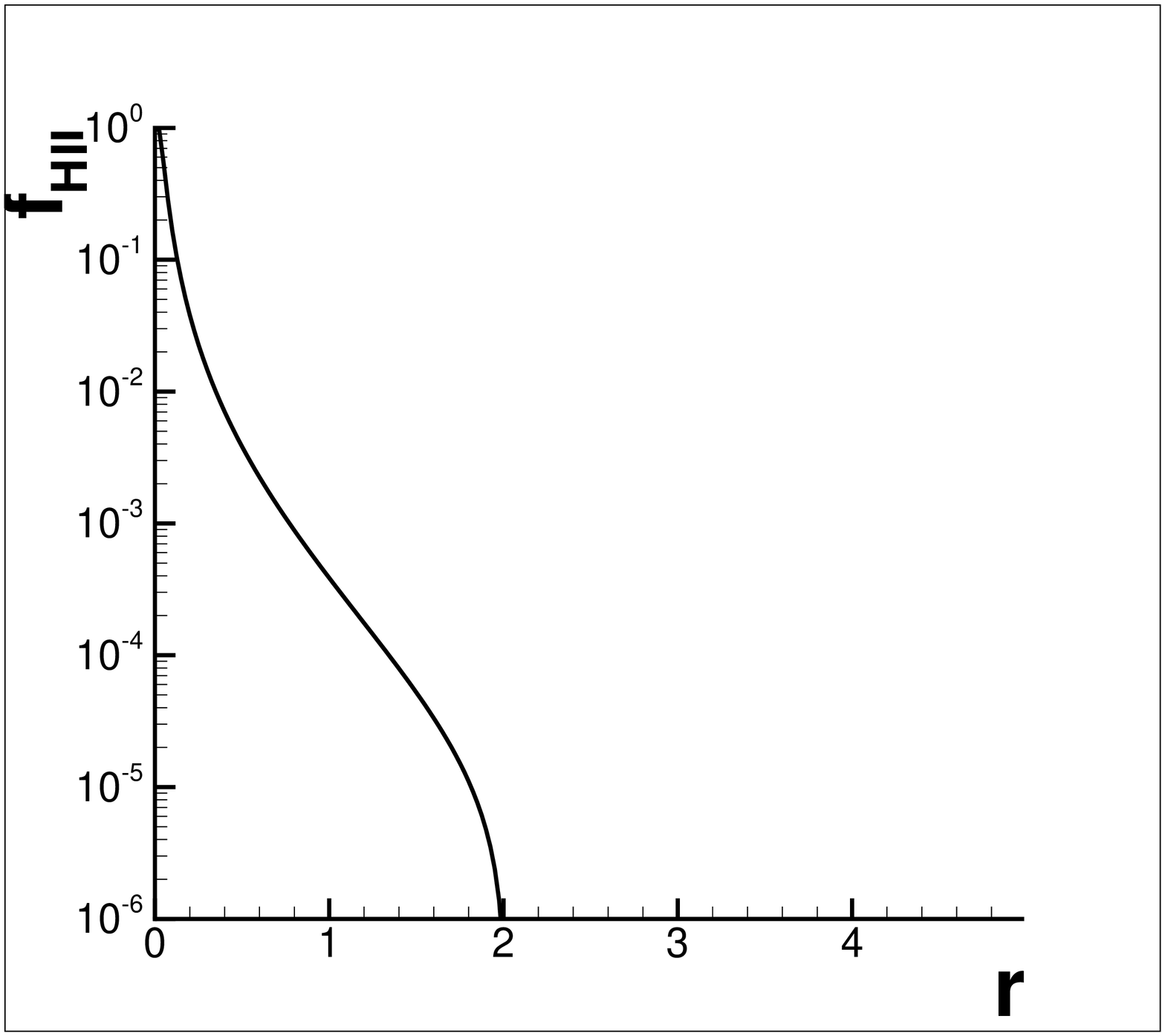}
}
\centerline{
\includegraphics[width=2.0in,height=1.5in]{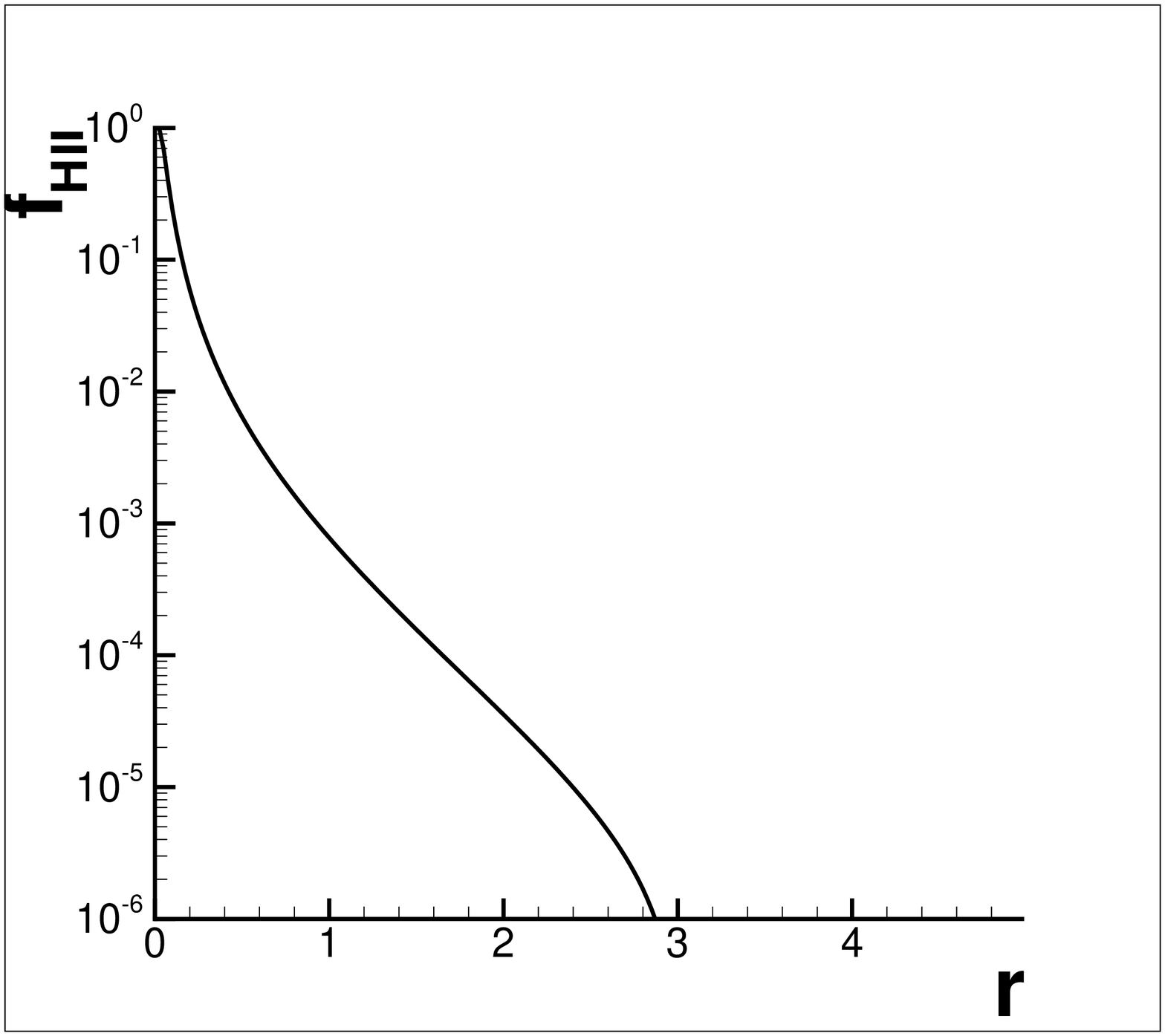}
\includegraphics[width=2.0in,height=1.5in]{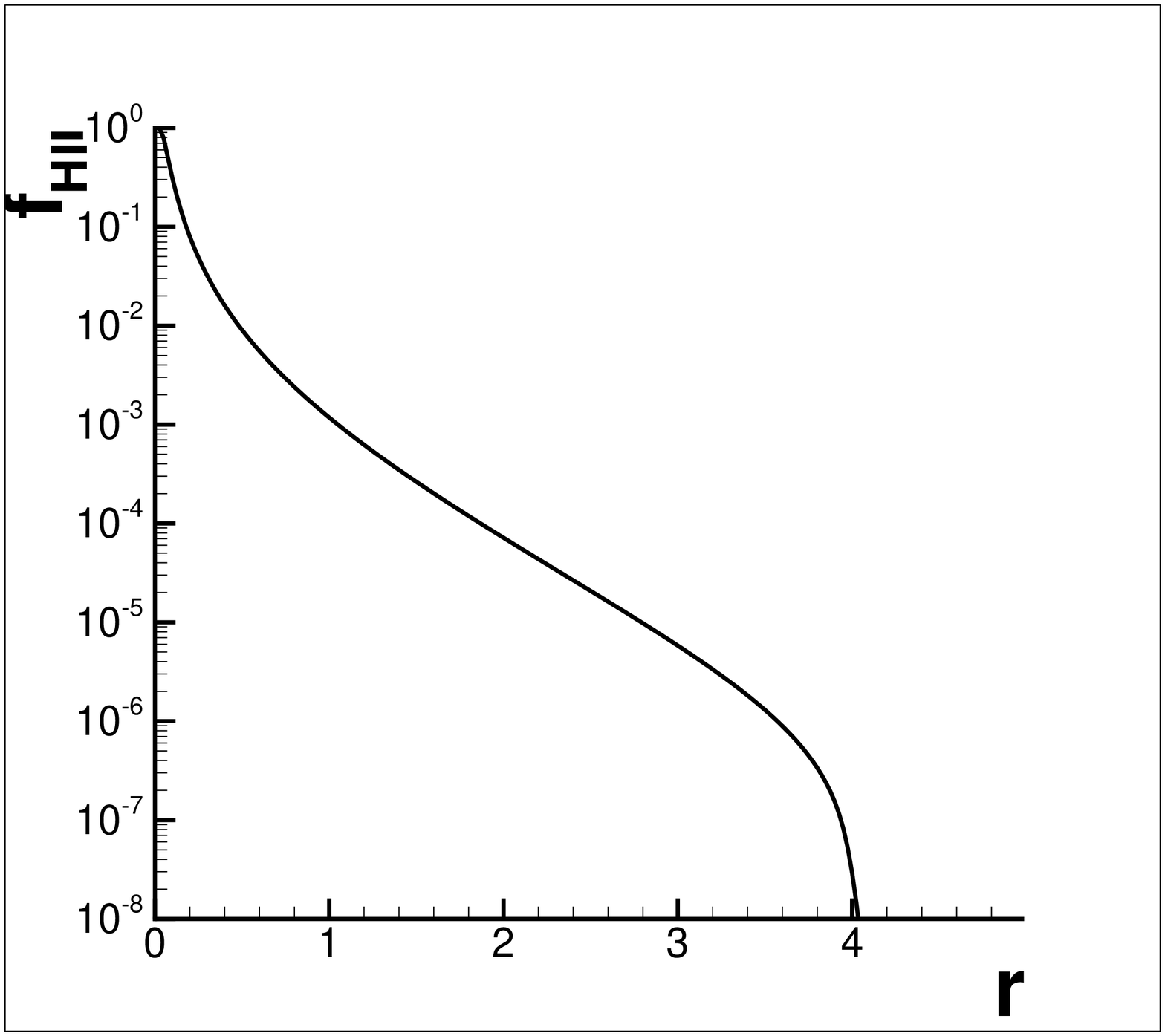}
}
\caption{$f_{\rm HII}(t,r)$ vs. $r$ at time $t=1.0$  (top
left), 2.0 (top right), 3.0 (bottom left), 4.0 (bottom right)
for a weak source $J''_0=0.001$ or $\dot{N}=5.05\times
10^{49}(1+z_r)^{-3}$. The parameter $N_r$ of the numerical calculation
is taken to be 200.}
\label{fig4}
\end{figure}

A solution numerical solution of the profile
$f_{\rm HII}(t,r)=1-f_{\rm HI}(t,r)$ at $t$=1, 2, 3, and 4 are shown 
in Figure \ref{fig4}, in which the source intensity is $J''_0=0.001$ or
$\dot{N}=5.05\times 10^{49}(1+z_r)^{-3}$ s$^{-1}$. The calculation 
is performed
with $N_r=200$.

We can see from Fig. 4 that the profile is very different from
strong source. The spatial range of $f_{\rm HI}(t,r)$ significantly 
less than 1 (or $f_{\rm HII}(t,r)$ significantly larger than zero) is 
much less than $t$, because the total number of photons 
emitted from sources within time $t$ is less than the number of atoms 
within
radius $r=t$. $f_{\rm HI}(t, r)$ is gradually increasing with $r$.
Although the non-zero range of the ionized fraction $f_{\rm
HII}=1-f_{\rm HI}(t,r)$ increases with time, the range $f_{\rm
HI}(t,r)>10^{-2}$ is always small. It implies that the ionized
sphere probably is not transparent to Ly$\alpha$ photons.

Figure \ref{fig4} shows that the effective ionized  radius $r_{\rm
HII}(t)$ [eq.(24)] for weak sources is $r_{\rm HII}(t)<t$, or
$r_{\rm HII}(t)/t<1$. The growth of the ionized range is much
slower than the radiative front. With the increase of source
intensity, $r_{\rm HII}(t)/t$ will gradually approach to 1, and
then, the solution will transfer to the case of strong sources.

\subsection{Evolution of the frequency spectrum}

Let us consider a point source emitting photons with a power law
frequency spectrum. The source function $S$ is then
\begin{equation}
\label{eq25}
S(t, {\bf x},{\nu}, {\bf n})=\left \{ \begin{array}{ll}
      (E(\nu)/V)\delta({\bf n - e_r}), & {\rm at \ center} \\
      0, & {\rm otherwise},
      \end{array}  \right .
\end{equation}
where $E(\nu)=E(\nu/\nu_0)^{-\alpha}$, $\alpha$ is the index of the
power law. Therefore, different from the problems of \S 3.1 and 3.2,
we should consider the variable $\nu$ of the frequency space, i.e.
$J$ is a function of $t$, $r$, and $\nu$. The equation of $J$ is
still the same as eq.(\ref{eq11}), but the absorption coefficient is
$\nu$-dependent
$$
k_{\nu} = \sigma(\nu)n_{{\rm
HI}}(t, {\bf x}),
$$
where the cross section $\sigma(\nu)=6.3\times
10^{-18}(\nu_0/\nu )^3\ {\rm cm^2}$. The boundary condition now
should be
\begin{equation}
\label{eq26} \lim_{r\rightarrow 0} J(t,r,\nu) = J_0
   (\nu/\nu_0)^{-\alpha}
\end{equation}
and the initial condition is
\begin{equation}
\label{eq27} J(t=0, r,\nu )=J_0(\nu/\nu_0)^{-\alpha}\theta(r=0)
\end{equation}

\begin{figure}[htb]
\centerline{
\includegraphics[width=2.0in]{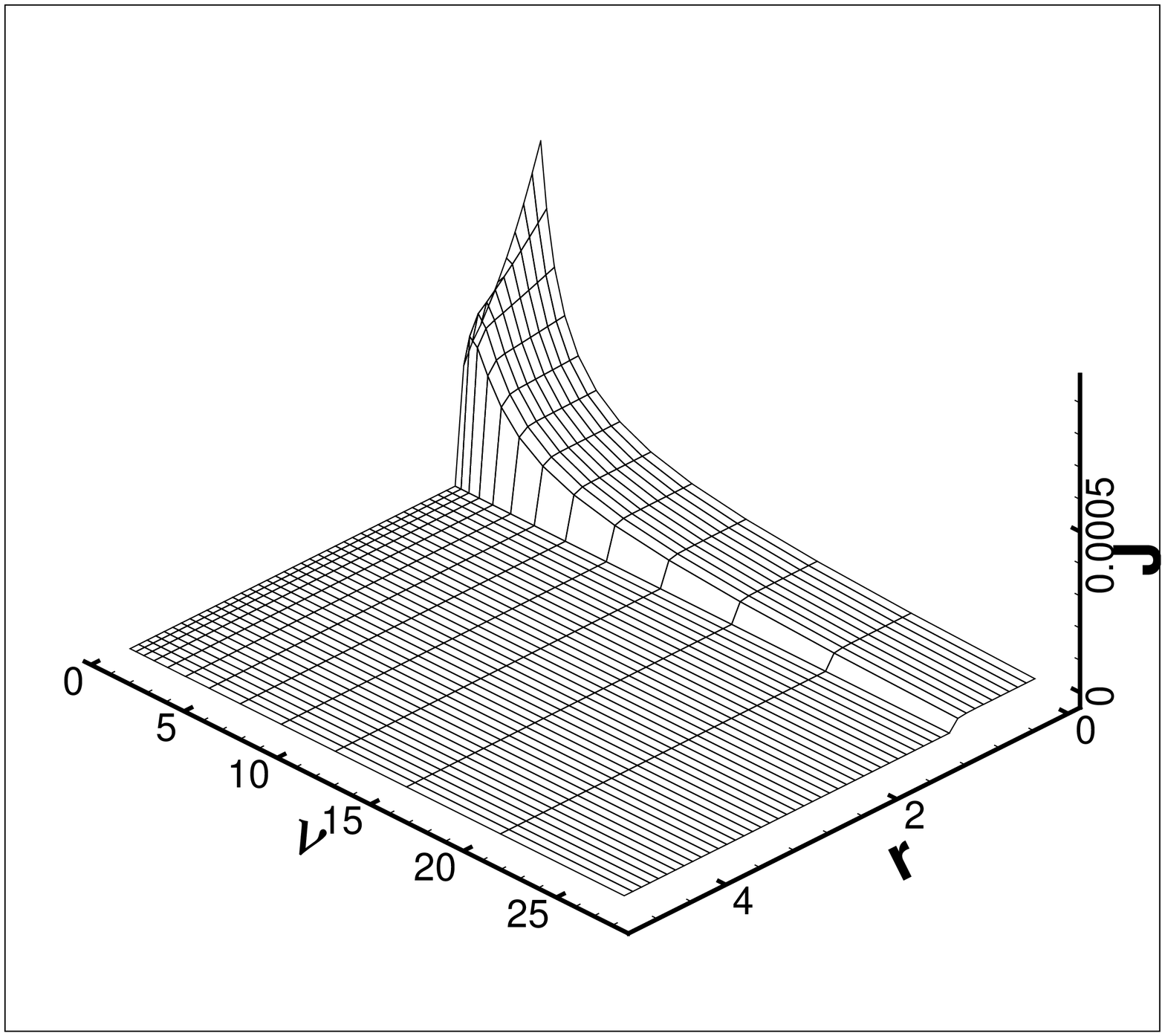}
\includegraphics[width=2.0in]{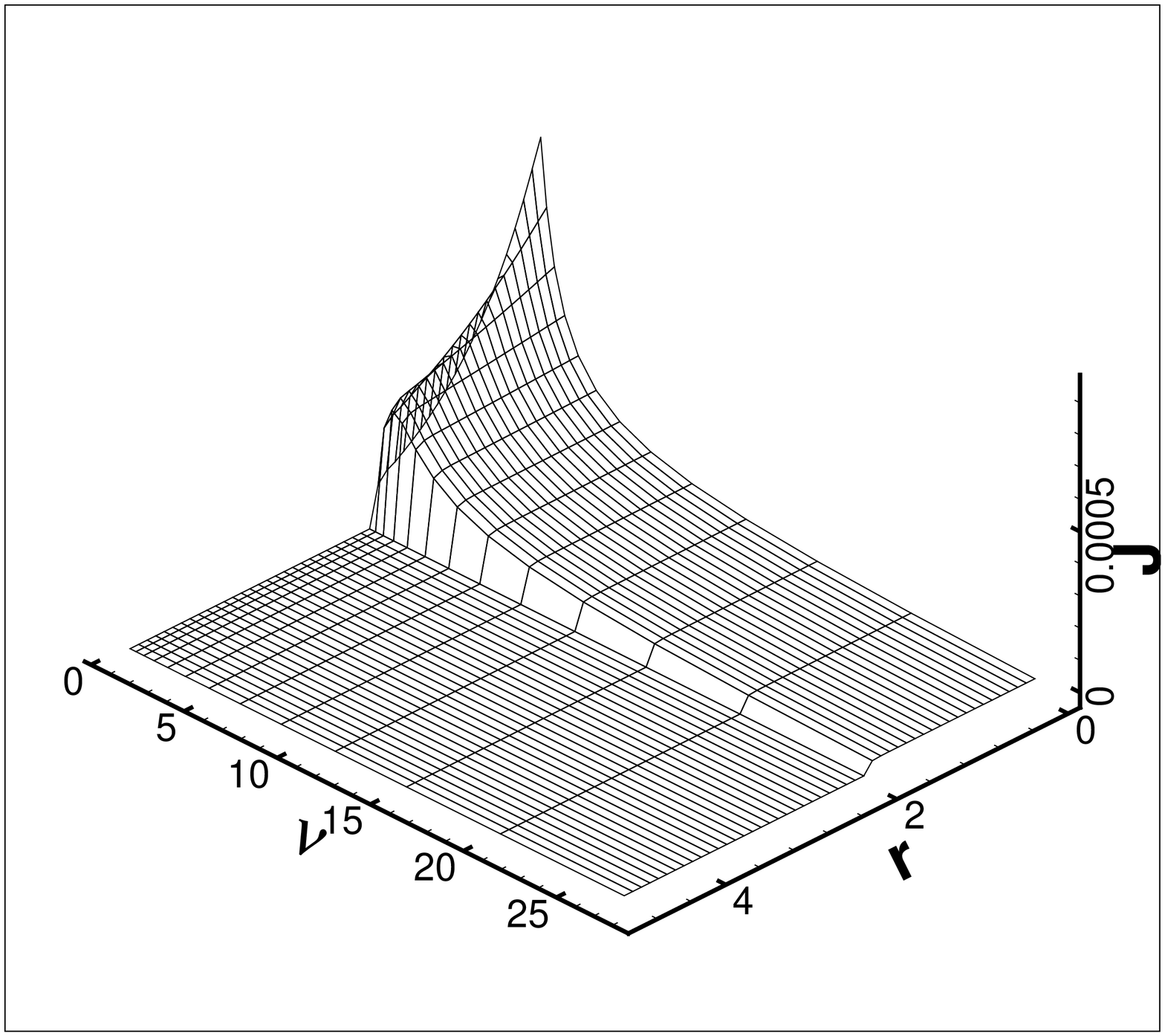}
} \centerline{
\includegraphics[width=2.0in]{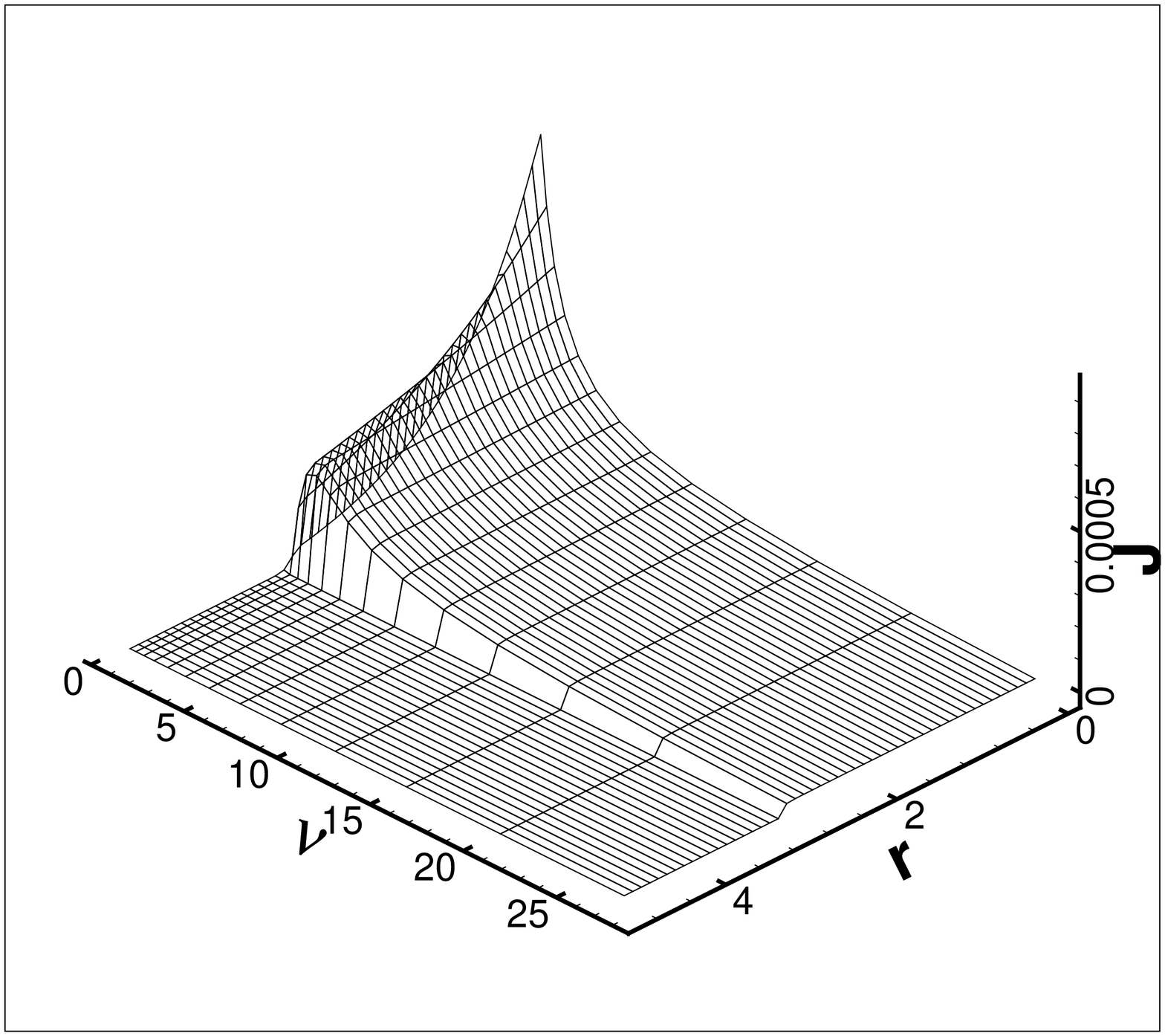}
\includegraphics[width=2.0in]{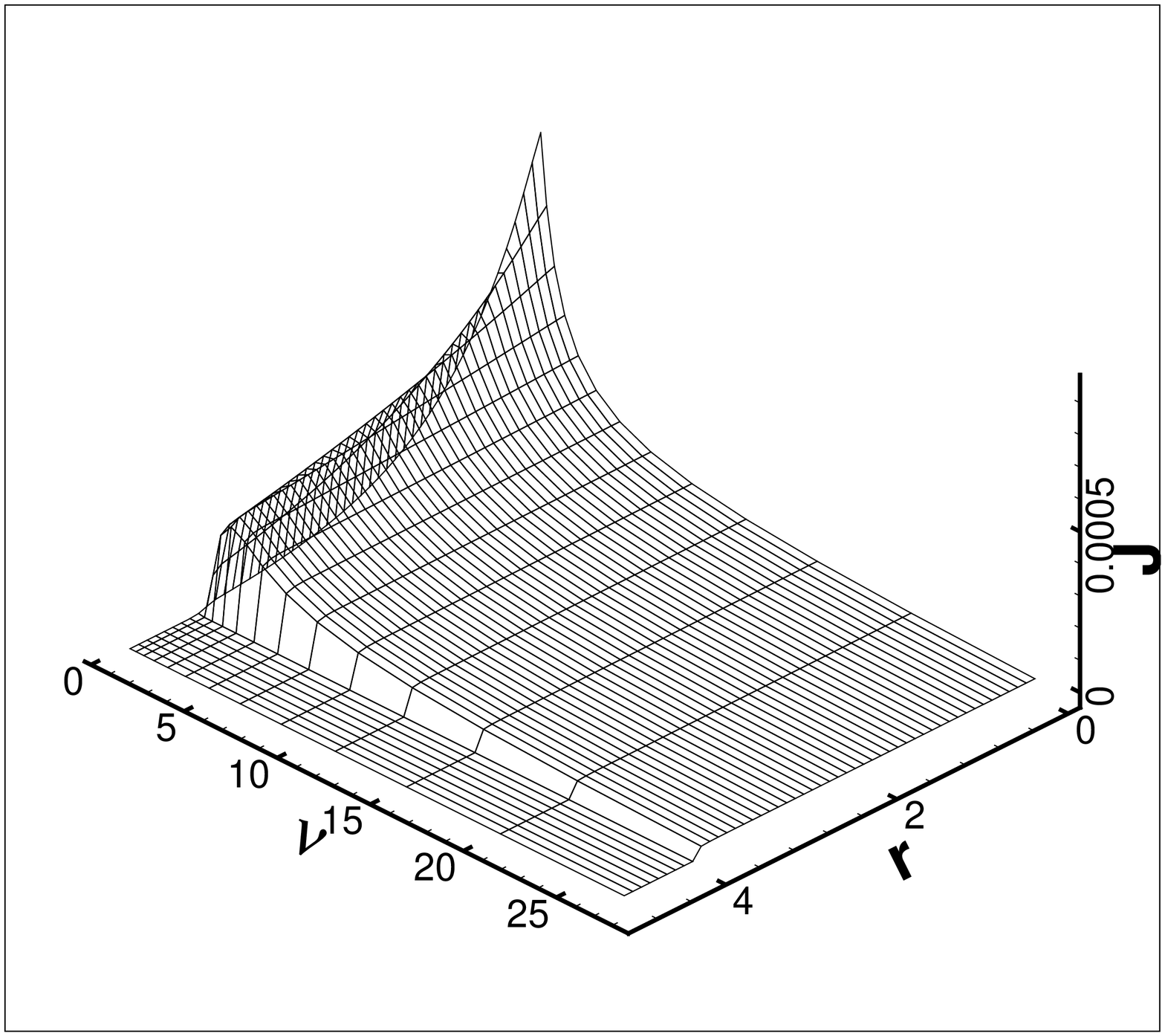}
} \caption{
A numerical result of the intensity $J''(t,r,\nu)$ for
$J''_0=0.001$ at time $t=1.0$ (top left), 2.0 (top right), 3.0
(bottom left), and 4.0 (bottom right). The mesh is taken to be
$N_r$=100, $N_\nu$=100, and $\nu_{\max}$=$10^6$.
}
\label{fig5}
\end{figure}

We still rescale the variables by $t'=cn \sigma(\nu_0) t$, $r'=n
\sigma(\nu_0) r $,  $\nu'=\nu/\nu_0$ and $J'' = J n
\sigma^2(\nu_0)/ch$. Then eq. (\ref{eq19}) is updated as follow,
\begin{equation}
\label{eq28} {{\partial J''}\over{\partial t'}}+{{\partial
J''}\over{\partial r'}}
   =- \left(\frac{1}{\nu}\right )^3 {f_{HI}} J''
\end{equation}
If we ignore the ionizing heating, $f_{\rm HI}$ or $n_{{\rm HI}}(t,
{\bf x})$ is still determined by eq.(\ref{eq18}), but $\Gamma_{\rm \gamma
HI}(t, {\bf x})$ should be given by
\begin{equation}
\label{eq29} \Gamma_{\rm \gamma HI}(t, {\bf x})= \frac{1}{r^2}
\int_{\nu_0}^{\infty}d\nu \frac{J(t,r, \nu)}{h\nu}\sigma(\nu).
\end{equation}
This integration could be worked out with a quadrature formula
accurate of order 4
\begin{equation}
\label{eq30} {\int_{\nu_0}^{\infty}}f(x)dx = \Delta x
{\sum_{j=j_0}^{\infty}}{w_j}f(j \Delta x) + O( \Delta x^4)
\end{equation}
where $\nu_0 = j_0 \Delta x$, and the weights $w_j$ are given by
$$
{w_{j_0}}=\frac{3}{8},  \quad {w_{j_0+1}}=\frac{7}{6}, \quad {w_{j_0+2}}
=\frac{23}{24},
$$
and
$$ {
w_{j_0+j}}=1, \qquad {\rm for} \ \  j > 2.
$$
By having a non-uniform mesh in the $\nu$-direction, one cannot use
this quadrature formula directly.  However, since a uniform mesh
size is used on $\xi$, with $\nu_j = 2^{\xi_j}$, we can perform the
numerical integration with respect to $\xi$ using (\ref{eq29}).

\begin{figure}[htb]
\centerline{
\includegraphics[width=2.0in]{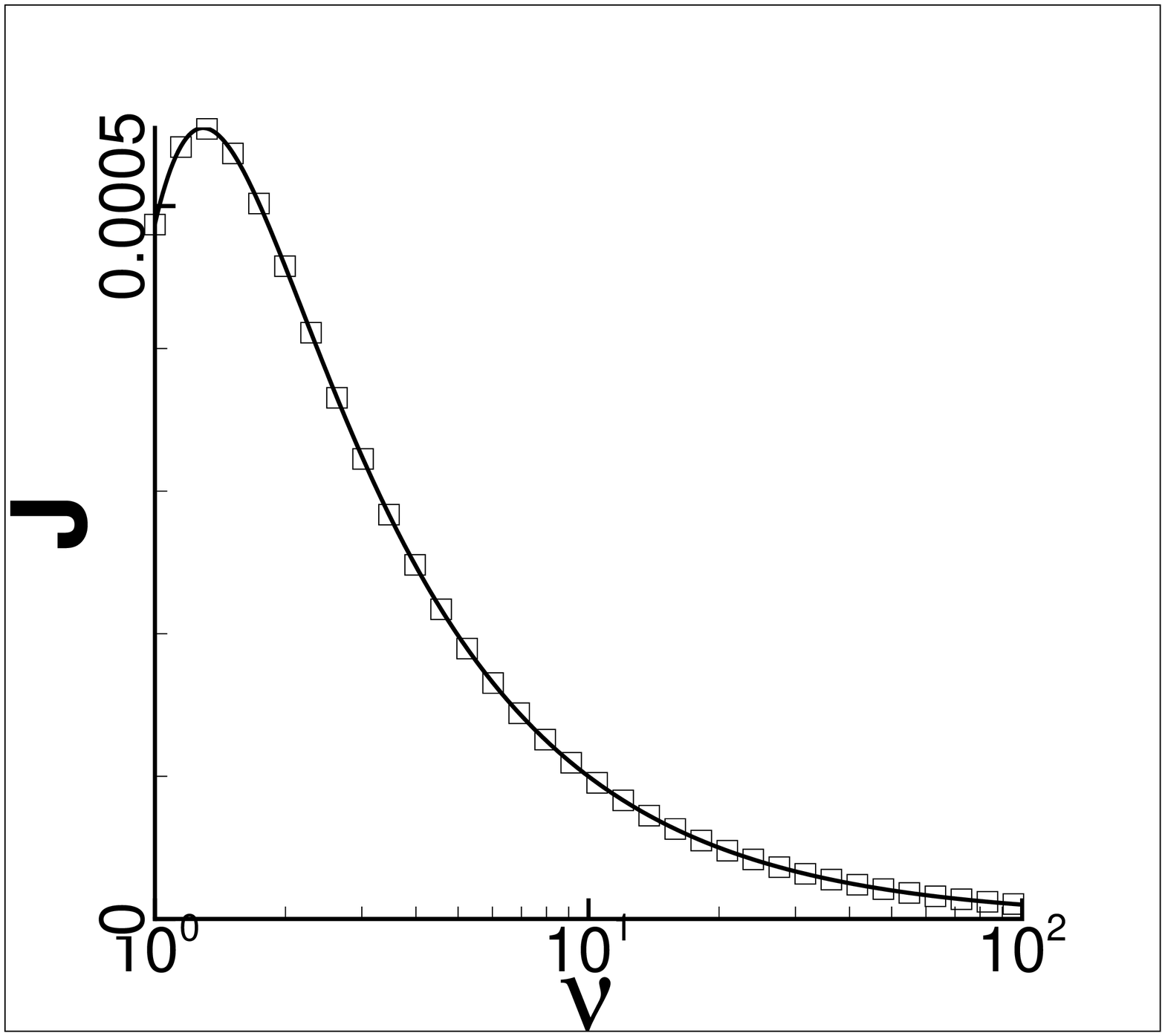}
\includegraphics[width=2.0in]{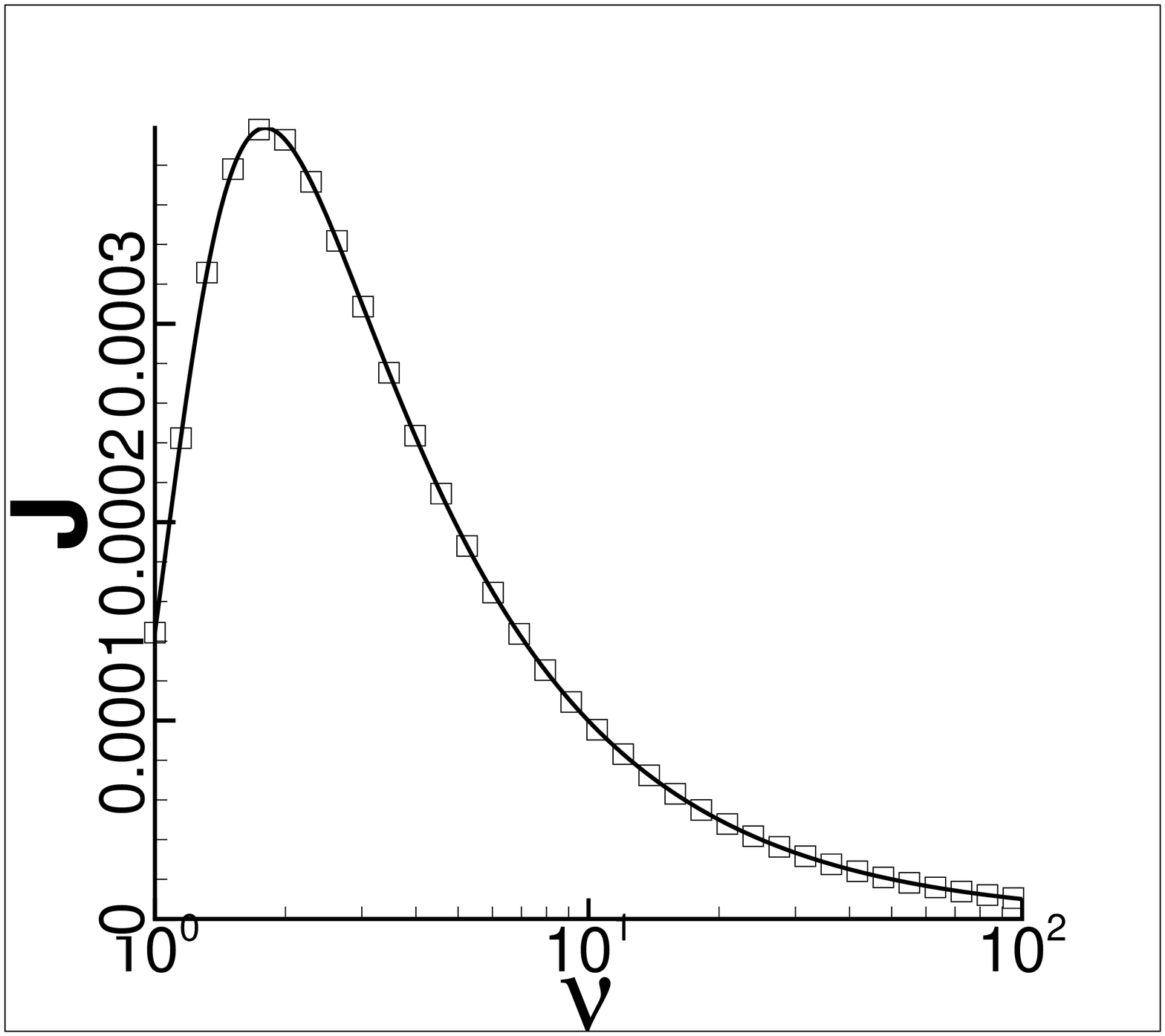}
} \centerline{
\includegraphics[width=2.0in]{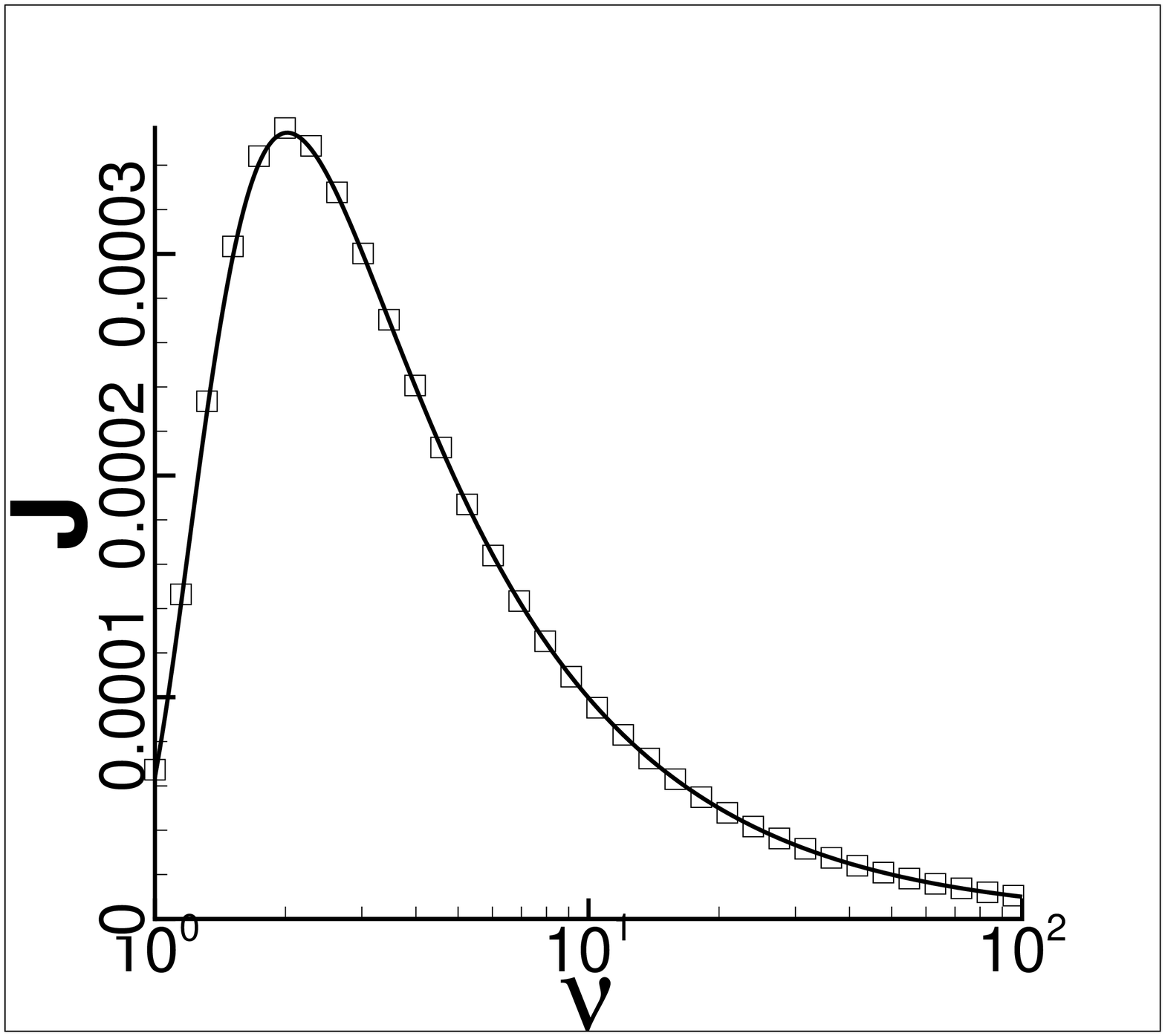}
\includegraphics[width=2.0in]{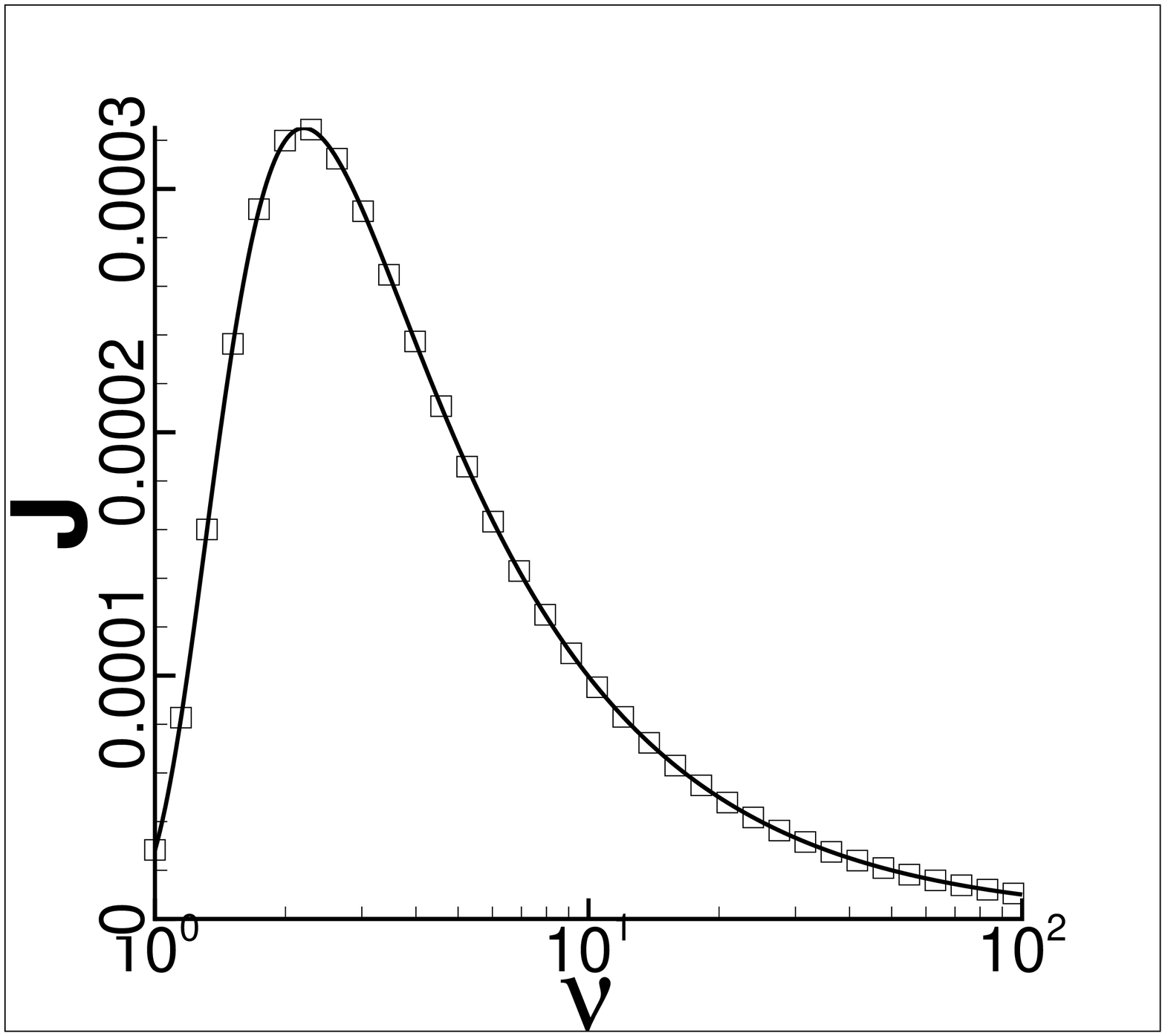}
} \caption{
A numerical solution of $J$ vs. $\log \nu$
 with $N_r=100$, $N_\nu=100$ (square) and the reference
 solution (solid) obtained with $N_r=1000$,
$N_\nu$=1000, at time $t=4.0$ and $r$=0.8 (top left),
 $r$=2.0 (top right), $r$=2.8(bottom left) and $r$=3.6 (bottom right).
}
\label{fig6}
\end{figure}

A numerical result of $J''(t,r,\nu)$ for $J''_0=10^{-3}$ and
$\alpha=1$ is shown in Figure \ref{fig5}, which is obtained by the
mesh $N_x$=100, $N_\nu$=100 and $\nu_{\max}$=$10^6$. It displays
the evolution of the frequency-dependence of $J''$. We notice that
for the high frequency $\nu>50$, the intensity $J''$ is almost
independent of $t$ and $r$, while the frequency spectrum at $\nu <
10$ is significantly dependent on both $t$ and $r$.

\begin{figure}[htb]
\centerline{
\includegraphics[width=2.0in]{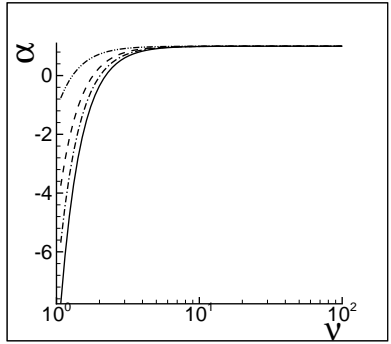}
}
\caption{
$\alpha$ vs. $\log \nu$ of the numerical solution
 with $N_r=100$, $N_\nu=100$ at time $t$=4.0 and $r$=0.8
(dash dot dot line), $r$=2.0 (dash line), $r$=2.8 (dash dot line)
and $r=$3.6 (solid line).
} \label{fig7}
\end{figure}

The evolution of the frequency spectrum can be seen more clearly in
Figure \ref{fig6}, which shows $J$ vs. $\nu$ at time $t=4.0$ and
position $r=0.8$, 2.0, 2.8 and 3.6. The results of Figure \ref{fig6}
given by $N_r=100$ and $N_\nu=100$ are the same as that obtained
with $N_r=1000$ and
$N_\nu=1000$. Therefore, the algorithm is also stable and
highly accurate in the frequency
space. From Figure \ref{fig6}, we can see a significant $r$-dependence of
the spectrum. At small $r$, the spectrum essentially is of
power law, while at $r=3.6$ it is similar to a spectrum of
self-absorption, i.e. it is low at $\nu \simeq 1$, and shows a peak
at $\nu \simeq 2$.

\begin{figure}
\centerline{
\includegraphics[width=2.0in]{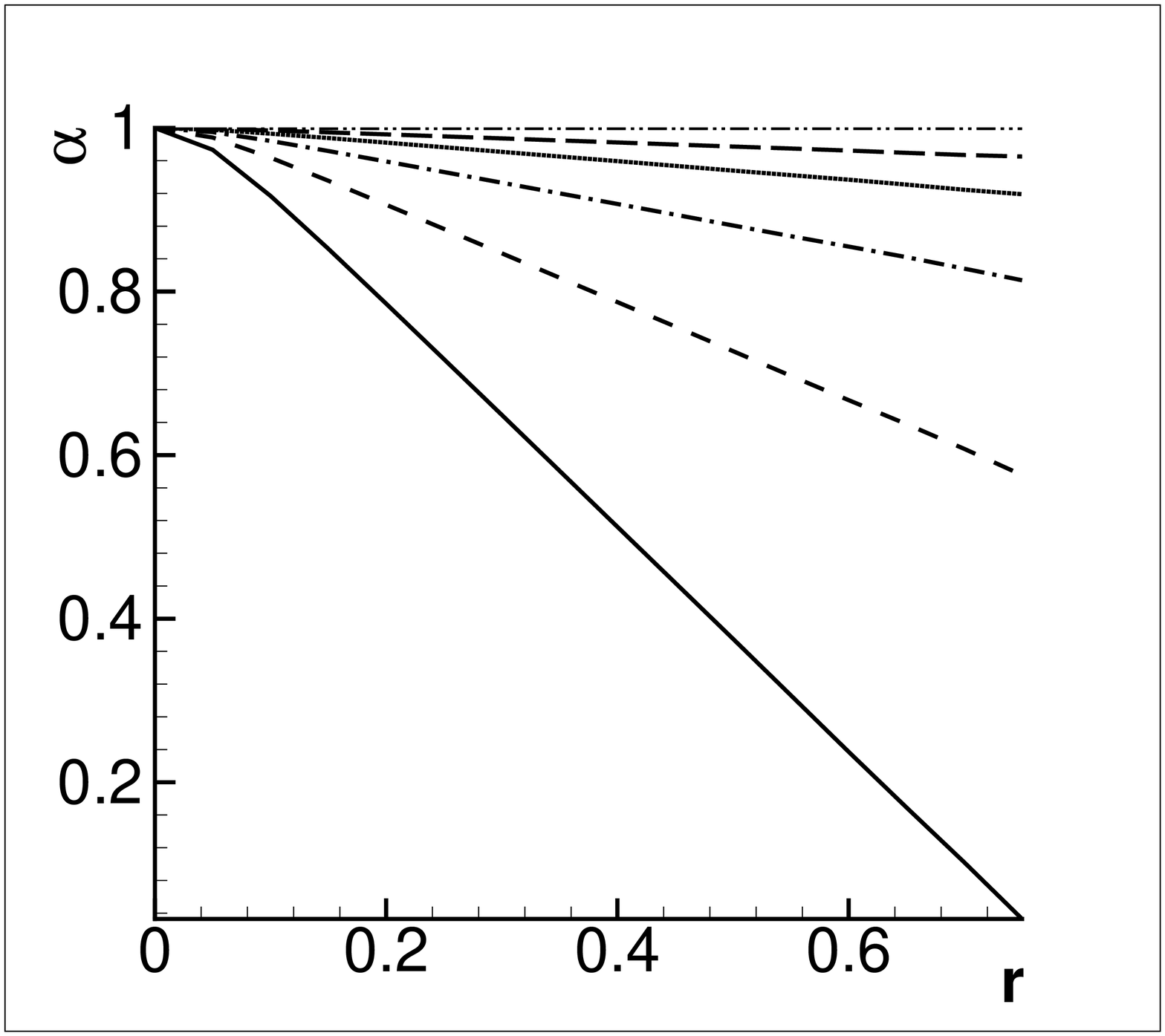}
\includegraphics[width=2.0in]{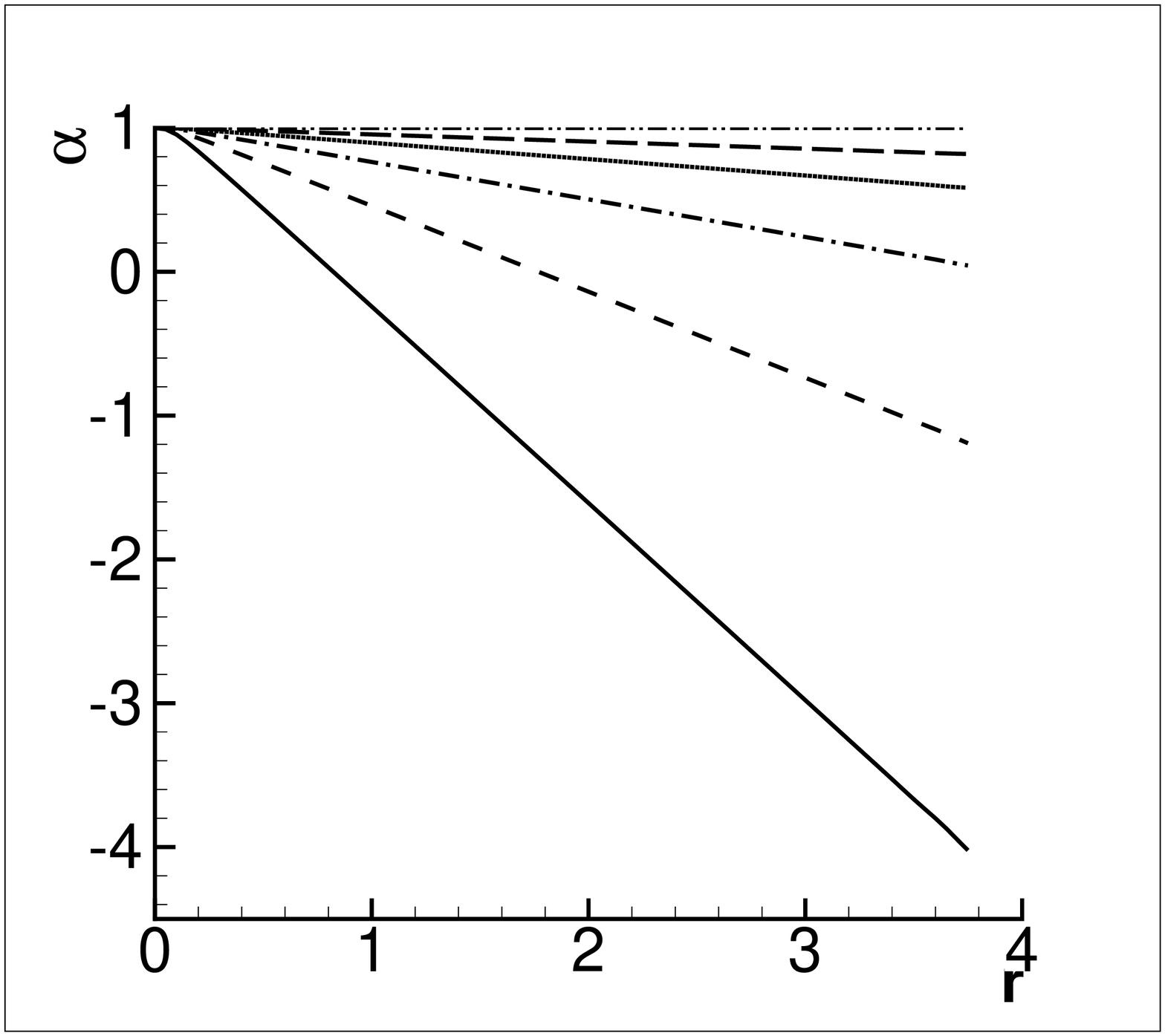}
} \caption{
$\alpha$ vs. $r$ with samples of $N_r=100$,
$N_\nu=100$ at time $t=1.0$ (left) and 4.0 (right) for wave bands
$\nu=1.32$ (solid line), $\nu=1.74$ (dash line), $\nu=2.29$ (dash dot
line), $\nu=3.02$ (dot line), $\nu=3.98$ (long dash line) and
$\nu=63.10$ (dash dot dot line).
}
\label{fig8}
\end{figure}

Generally, self-absorption leads to the lack of photons with $\nu
\simeq 1$, and therefore, to the hardening of the frequency
spectrum of photons. One can measure the hardening by the index
of power law defined by
\begin{equation}
\label{eq31}
\alpha=-{\partial \ln J \over \partial\ln \nu}.
\end{equation}
Figure \ref{fig7} plots $\alpha$ vs. $\nu$ at different positions
$r$ and shows that $\alpha$ becomes smaller at $\nu < 10$. Figure
\ref{fig8} gives $\alpha$ vs. $r$ for several $\nu$. It shows that
$\alpha$ becomes smaller at larger $r$.

\section{Concluding remarks}

We described a numerical solver for radiative transfer problems
based on the WENO scheme modified with anti-diffusive flux
corrections. It has high order of accuracy and good convergence, and
is also highly stable and robust for solving problems with both
discontinuities and smooth solution structures. Using the
Str\"omgren sphere or ionized sphere as numerical tests, we showed
that this code is able to resolve the sharp ionized front under a
wide parameter range, such as intensity of the source $\dot{N}$
varying from $10^{49}-10^{57}(1+z_r)^{-3}$ s$^{-1}$, which covers
various sources responsible for the early reionization in the universe.

Since the WENO scheme needs more floating point operations per cell
than those of the PPM and TVD schemes, it leads to twice or more
loss of computational speed. Nevertheless, it has been already
successfully applied to kinetic equations of distribution function
in phase space with one or two spatial dimensions and two or three
phase space dimensions. It can provide the solution of evolution in
both the physical space as well as the frequency space. Therefore,
the WENO scheme would be useful to study the early stage of
reionization, in which ionized region is generally an isolated patch
given by individual source with various configuration.

We demonstrate this algorithm with the problems of spherical ionized
region exposed to a point source. It is shown that the radial
profile of the fraction of ionized hydrogen is sensitively dependent
on the intensity of the sources. Although for strong sources the
Str\"omgren sphere provides a good approximation to the ionized
sphere, the profile of the sphere actually can not always be
approximated as fully ionized spheres sharply separated with neutral
hydrogen region. Even for strong sources, the neutral hydrogen
remained in the Str\"omgren ionized sphere may still be significant
and important. There are 1-D radiative transfer codes, such as
CLOUDY94 (e.g. Maselli et al. 2003), which are capable of capturing the sharp
boundary of the Str\"omgren sphere. However, they yield large errors of
the neutral hydrogen remained in the sphere. Our result shows that
the WENO codes can effectively handle the uncertainty around
discontinuities.

We studied the time-dependence of the ionized radius $r_{\rm
HII}(t)$. We found that for strong sources, $r_{\rm HII}(t)\simeq t$
when $r_{\rm HII}< R_s$. This is similar to the result of Madau \& Rees
2000. For weak sources, we have $r_{\rm
HII}(t) < t$. However, the evolution of $r_{\rm HII}(t)$ can not be
fitted by $r_{\rm HII}(t)\propto t^{1/3}$, which will yields unphysical
result $dr_{\rm HII}(t)/dt >1$ at small $t$ (e.g. Cen \& Haiman 2000;
Yu \& Lu 2005). Moreover, the WENO algorithm provides also the
solution of the evolution of the photon's frequency spectrum of an
ionized sphere.

The WENO scheme for radiative transfer problems could be
incorporated with the Euler hydrodynamics. For instance, it is
straightforward to generalize the calculation of a Str\"omgren
sphere to take account of the hydrodynamic effects of the hydrogen
gas sphere even when the gaseous sphere undergoes shocks.

\noindent{\bf Acknowledgments}

This work is supported in part by the US NSF under the grants
AST-0506734 and AST-0507340. LLF acknowledges support from the
National Science Foundation of China under the grant 10573036.

\appendix

\section{Radiative transfer equations}

Considering a universe described by a Robertson-Walker metric, we have
\begin{equation}
\label{eqA1} ds^2=
g_{\mu\nu}dx^{\mu}dx^{\nu}=dt^2-a^2(t)g_{ij}x^ix^j,
\end{equation}
where we take $c=1$. The summation is over  0 to 3 for repeated
Greek indices, 0 to 3 for Latin indices, $a(t)$ is the cosmic
factor and $g_{ij}=\delta_{ij}-h_{ij}$. We consider below the flat
universe, i.e. $h_{ij}=0$. The distribution function $f(t, x^{i},
p^{\alpha})$ of photons is defined by
\begin{equation}
\label{eqA2} dN = -f(x^{\alpha},
p^{\alpha})2\delta^D(p^{\alpha}p_{\alpha})
d\mathcal{V}_xd\mathcal{V}_p,
\end{equation}
where $dN$ is the number of photons in the invariant phase space
volume element of
\begin{equation}
\label{eqA3} d\mathcal{V}_x=a^3p^0dx^1dx^2dx^3 \hspace{1cm}
d\mathcal{V}_p=dp_0dp_1dp_2dp_3.
\end{equation}
 The variables $x^i$ is comoving coordinate,
$p^0=dt/d\lambda$, and $p^i=a(t)dx^i/d\lambda$ are 4 momentum,
$\lambda$ being the affine parameter. The Dirac-delta function
$\delta^D(p^{\alpha}p_{\alpha})$ in eq.(\ref{eqA2}) is due to the 4
momentum $p^{\alpha}$ of photons to be null, i.e.
$p_{\alpha}p^{\alpha}=0$.

The Boltzmann equation of the distribution function $f(x^{\alpha},
p^{\alpha})$ is given by (e.g. Bernstein 1988)
\begin{equation}
\label{eqA4} \frac{\partial f}{\partial t}+ \frac{d
x^i}{dt}\frac{\partial f}{\partial x^i}
-\frac{\dot{a}}{a}p^i\frac{\partial f}{\partial p^i} =Q,
\end{equation}
where $Q$ is collision term, of which the variables are also
$t,x^i,\nu,n^i$. Define $n^i=a(dx^i/dt)$, it is a unit vector in
the direction of photon propagation, ${\bf n\cdot n}=1$. With
$n^i$, we have $p^i= p^0n^i$, or $p^0=\hbar \nu/c$, the frequency
$\nu$ of photon. From eq.(\ref{eqA4}), the equation of the
distribution function $f(t,x^i,\nu, n^i)$ is
\begin{equation}
\label{eqA5} \frac{\partial f}{\partial t}+\frac{1}{a}
\frac{\partial }{\partial x^i}(n^if) -H\nu\frac{\partial
f}{\partial \nu} = Q,
\end{equation}
where the Hubble constant $H=\dot{a}/a$. From the definition
eq.(\ref{eqA2}) of $f$, the specific intensity $J\propto \nu^3 f$,
which is also a function of $t$, ${\bf x}$, $\nu$ and $n^i$.
Moreover, if $Q$ is mainly given by absorption and emission
sources, Eq.(\ref{eqA5}) yields the equation of specific intensity
as follows
\begin{equation}
\label{eqA6}
        {\partial J\over\partial t} +\frac{1}{a}
        {\partial \over\partial x^i}(n^i J) -
        H\left(\nu{\partial J\over\partial\nu}-3J\right) =
        - k_\nu J + S.
\end{equation}
where $k_\nu$ is the absorption coefficient, and $S$ the source
function. Eq.(\ref{eqA6}) is the equation of radiative transfer in
the universe. The equation (\ref{eqA6}) can be rewritten as in the
format of a conservation flux as
\begin{equation}
\label{eqA7}
        {\partial J\over\partial t} +
        {\partial \over\partial x^i}\left (\frac{n^i}{a} J \right )
        +
        {\partial \over\partial \omega}(HJ) =
        - (k_\nu+3H) J + S,
\end{equation}
where $\omega\equiv \ln 1/\nu$. Therefore, it can be solved by the
WENO scheme.

\end{document}